\documentclass[11pt,a4paper]{article}
\usepackage{amsmath,amsthm,amssymb}
\usepackage[titletoc,toc,title]{appendix}
\usepackage{bbm}
\usepackage[latin1]{inputenc}
\usepackage[english]{babel}
\usepackage{csquotes}
\usepackage{pgfplots}
\usepackage{pgfplotstable}
\usepackage[authoryear]{natbib}
\pgfplotsset{
    compat=1.3,
    xticklabel={$\mathsf{\pgfmathprintnumber{\tick}}$},
    scaled ticks=false,
    every axis/.append style={
        font=\footnotesize,
        width=6.5cm,
        height=4.9cm,
        ylabel shift=-2pt, 
    }
}

\usepackage{graphicx}
\usepackage{float}
\restylefloat{table}
\usepackage[hidelinks]{hyperref}
\usepackage[
       paper=a4paper,
       tmargin=3cm,           
       bmargin=3cm,           
       lmargin=2.5cm,             
       rmargin=3cm]{geometry}   

\def \Sum{\displaystyle\sum}

\def\<{\left<}
\def\>{\right>}
\def\({\left(}
\def\){\right)}

\def\9{\infty}

\def\N{\mathbb N}

\def\E{\mathbb E}
\def\P{\mathbb P}
\def\Q{\mathbb Q}

\def\shl{{\cal L}}

\newtheorem{theorem}{Theorem}

\newtheorem{remark}[theorem]{Remark}

\newcommand{\beqnar}{\begin{eqnarray*}}
\newcommand{\eeqnar}{\end{eqnarray*}}
\newcommand{\ba}{\begin{array}}
\newcommand{\ea}{\end{array}}

\begin{document}

\title{Gas storage valuation and hedging. \\ A quantification of the model risk.}
\author{ Patrick Henaff (1), Ismail Laachir (2) and Francesco Russo (3)}
\date{December 12th 2013}
\maketitle
\vspace{10mm}
{\bf Abstract.}

This paper focuses on the valuation and hedging of gas storage facilities, using a spot-based valuation framework coupled with a financial hedging strategy implemented with futures contracts.
The first novelty consist in proposing a model that unifies the dynamics of the futures curve and the spot price, which accounts for the main stylized facts of the US natural gas market, such as seasonality and presence of price spikes. The second aspect of the paper is related to the quantification of model uncertainty related to the spot dynamics.\\

{\bf Key words and phrases.}
Energy markets; commodities; natural gas storage; model uncertainty.

{\bf JEL Classification:} C4; C5; C6; C8; G11; G13; G17.

\medskip
{\bf MSC Classification 2010:} 91G10; 91G60; 91G70; 91G80.

\begin{itemize}
\item[(1)] Patrick Henaff, IAE Paris, Université Paris I - Panthéon Sorbonne, France

\item[(2)] Ismail Laachir, Université de Bretagne-Sud and ENSTA ParisTech, France
 
\item[(3)] Francesco Russo, ENSTA ParisTech, Unit\'e de Math\'ematiques
appliqu\'ees, France.
\end{itemize}

\section{Introduction}

\label{SIntro}

Natural gas storage units are used to reconcile the variable seasonal demand for gas with the more constant rate of natural gas production. These gas storage facilities are mainly owned by distribution companies which use them for system supply regulation, and to reduce the risk of shortages. In fact, due to regulation laws, local distribution companies are obliged to own storage units, to ensure their gas supply and to be able to meet any sudden increase in demand or any disruption in the pipeline transportation system.\\

In practice there are several techniques for gas storage valuation: 
The two most popular are the classical \textit{intrinsic valuation} based on physical futures contracts, which is equal to the optimal discounted value of calendar spreads, and the \textit{extrinsic valuation}, which uses spot trading strategies. Traditionally the demand for natural gas is  seasonal, with peaks during winter, and lows during summer. This motivates the first valuation methodology, which exploits the predictable seasonal shape of the natural gas futures curve.  Following this strategy, the storage manager observes the initial futures curve, at the beginning of the storage contract, decides to buy/sell multiple futures contracts, and consequently receives/delivers natural gas at their expiration. In order to determine the optimal futures positions, a linear optimization problem has to be solved, with constraints imposed by the physical and financial conditions of the storage contract (see Annexe \ref{appB} which refers to 
\citet{eydelandBook}). We emphasize that storage manager keeps the optimal futures positions for the whole storage contract duration, so this strategy does not take advantage of possible profitable movements of the futures curve. \\

This static methodology is extended by \citet{GrayIntrinsic} to the \textit{rolling intrinsic valuation}, to take advantage of the changing dynamics of the futures curve. In that case, optimal futures positions are chosen at the beginning of the storage contract, but when the futures curve moves away from its initial shape, the new optimal futures positions is recalculated and if managers find it more profitable, the portfolio is rebalanced. Hence, the evaluation of the rolling intrinsic strategy requires a model for the dynamics of the futures curve.\\

These two futures-based approaches capture the predictable seasonal pattern of natural gas prices: 
In particular they lead  to buy cheap summer futures and 
sell expensive winter futures, and the storage value obtained greatly depends on the seasonal spread between cold and warm periods of the year. \\

The intrinsic valuation methodology has been popular in the storage industry, especially during periods when seasonal patterns are very pronounced. However, during the last years, the seasonal spreads have been reduced, which puts into perspective the futures-based methodology. In fact, the \textit{2011 State of the Markets} report, by the US Federal Energy Regulatory Commission (FERC) \cite{FERCStateMarket}, noticed the following:
 \enquote{We have also seen a decline in the seasonal difference between winter and summer natural gas prices. Falling seasonal 
spreads reflect increased production and storage capacity, as well as greater year-round use of natural gas by power generators. This decline has developed over the past several years and we expect the trend to continue.} 
The narrowing winter/summer spread, mentioned earlier, is mainly due to two factors that put a downward pressure on winter gas prices and an upward pressure on summer prices. The first factor is the recent surge of the non conventional shale gas supply, with geographical locations that are closer to gas consumption areas. The main result of this new abundant source of gas is a downward pressure on winter prices. The second factor  is related to power consumption by cooling systems during summer periods and the growing use of natural gas as a fuel for electricity generation, instead of other sources of power. This puts an upward pressure 
on summer gas prices. \\

The combination of these two factors has the logical consequence of narrowing the seasonal spreads between winter and summer prices, diminishing  the intrinsic value of gas storage units. The use of futures contracts, exclusively as instruments to monetize the value of the gas storage, is no longer sufficient: in particular it sometimes fails to recover the operating expenses. This motivates 
the interest in  valuation strategies based on the gas spot prices 
(instead of futures). This so-called \textit{extrinsic valuation} can still take advantage of the remaining seasonality and more importantly allows the monetization of the high volatility of natural gas spot prices.\\

The first contribution of this paper to the topic is the proposal of
a new modeling framework, unifying the natural gas futures curve and the spot price, taking into account the stylized facts that are essential in 
the gas storage valuation problem i.e., seasonality and spikes.
The second aspect of the paper is related to the quantification of model
 uncertainty related to the spot dynamics.\\
 
In fact, the main result of  Section \ref{S7} is the significant sensitivity of gas storage value with respect to the specification and estimation of the spot model.
This result puts into perspective the extensive literature on gas storage valuation, and calls for a more careful study of the model risk inherent to our problem.  We believe that it is crucial to turn more attention to the choice of the spot-futures modeling framework, rather than concentrate all the effort on the specification of an optimal trading strategy.

This paper is organized as follows. In Section \ref{stylFact} we describe the important stylized facts 
related to natural gas markets. In Section \ref{S3} we describe
the characteristics of the gas storage unit and valuation using an optimal spot strategy, and the futures-based hedging methodology.
Section \ref{S4} is devoted to a review of the modeling
approaches in the gas storage literature. In Section \ref{S5}
we introduce the modeling framework combining futures and 
spot dynamics and in Section \ref{SNumRes} we perform 
several numerical tests. In Section \ref{S7}  we introduce 
two natural model risk measures to quantify the sensitivity
 of a class of models 
with respect to the parameters; those risk measures are computed
in several test cases.

\section{Natural gas stylized facts}
\label{stylFact}
In this section, we highlight important stylized facts about natural gas 
markets that influence the value of a storage unit. These properties are related to the demand and use of natural gas. In fact, the demand for natural gas for
 heating in cold periods of the year produces a seasonal behaviour
  for prices during winter periods, while unpredictable changes
 in weather can cause sudden shifts in gas prices. These facts are the 
two main sources of value for a gas storage unit, which allows for the exploitation of seasonality and sudden variations in demand.\\

As for all other commodities, the prices of natural gas (NG)  
are influenced by their geographical location.
 In our study, we will be interested in  the 
United States market, specifically in  a storage location near Henry Hub (Louisiana), which justifies the use of gas daily spot prices and the Nymex natural gas futures as hedge instruments.
Essentially, it is possible to buy natural gas using spot or futures contracts. In the spot market the prices are settled every day for a delivery on the day after.
In the futures market the prices of 72 monthly futures contracts are available on every business day, but there are only about 24 or fewer liquid contracts.\\

In what follows, $S_t$ will denote the spot price of natural gas at date $t$, and $(F(t,T_i))_i$ represent the futures contracts prices at $t$, for a set $\{T_i\}$ of maturities. We consider monthly spaced maturities, so every futures contract is related to a month of a year. Also, we denote by $P_t$ the price of prompt contract, i.e. the futures contract with the closest maturity to current time $t$. Natural gas prices are quoted in U.S. dollars per million British thermal units (MMBtu).\\

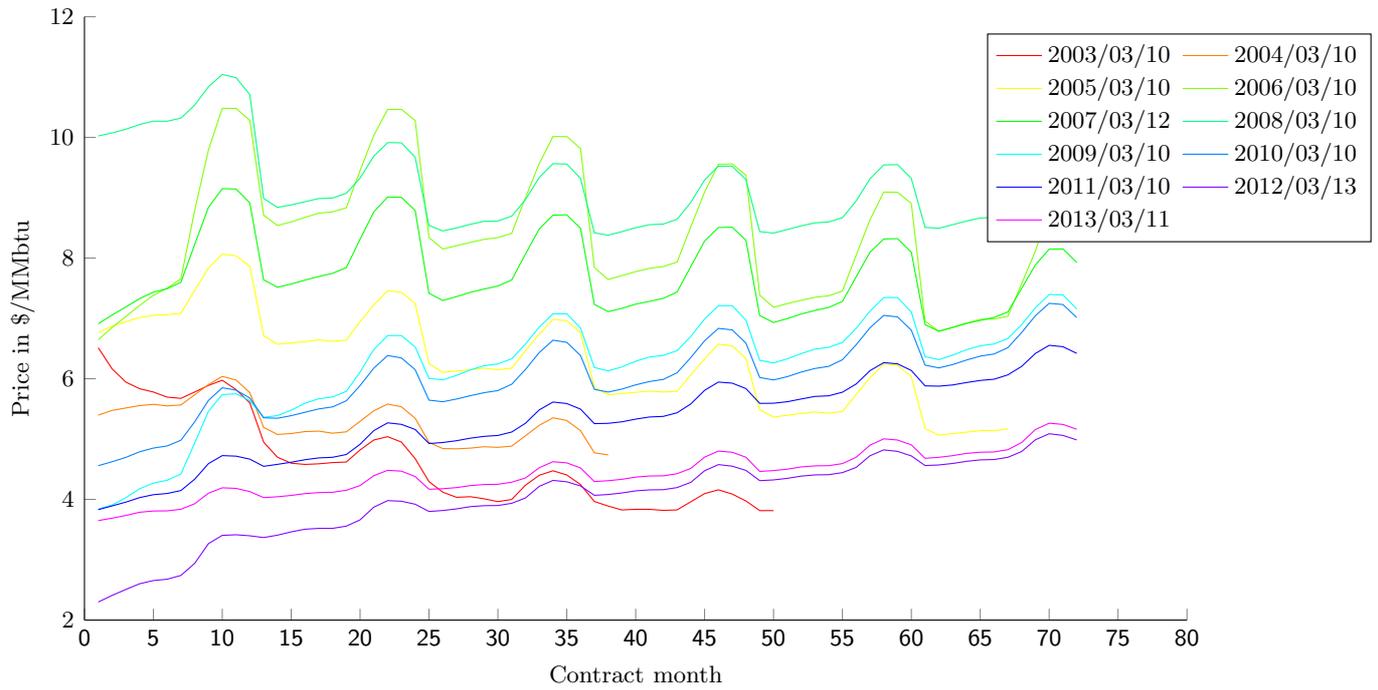
\begin{figure} 
\newlength\figureheight 
\newlength\figurewidth 
\setlength\figureheight{8cm} 
\setlength\figurewidth{14.5cm}
%
%
%
%

\definecolor{mycolor1}{rgb}{1,1,0}%
\definecolor{mycolor2}{rgb}{0.5,1,0}%
\definecolor{mycolor3}{rgb}{0,1,0.5}%
\definecolor{mycolor4}{rgb}{0,1,1}%
\definecolor{mycolor5}{rgb}{0,0.5,1}%
\definecolor{mycolor6}{rgb}{0.5,0,1}%
\definecolor{mycolor7}{rgb}{1,0,1}%

\begin{tikzpicture}

\begin{axis}[%
width=\figurewidth,
height=\figureheight,
scale only axis,
xmin=0,
xmax=80,
xlabel={Contract month},
ymin=2,
ymax=12,
ylabel={Price in \$/MMbtu},
axis x line*=bottom,
axis y line*=left,
legend columns=2, 
legend style={at={(0.818692044899951,0.626278118609407)},anchor=south west,draw=black,fill=white,legend cell align=left, /tikz/column 2/.style={
                column sep=2pt,
            }}
]
\addplot [
color=red,
solid
]
table[row sep=crcr]{
1 6.515\\
2 6.165\\
3 5.945\\
4 5.835\\
5 5.78\\
6 5.695\\
7 5.675\\
8 5.78\\
9 5.89\\
10 5.975\\
11 5.82\\
12 5.6\\
13 4.95\\
14 4.7\\
15 4.6\\
16 4.58\\
17 4.59\\
18 4.61\\
19 4.62\\
20 4.82\\
21 4.985\\
22 5.04\\
23 4.95\\
24 4.675\\
25 4.295\\
26 4.12\\
27 4.035\\
28 4.045\\
29 4.01\\
30 3.963\\
31 3.999\\
32 4.235\\
33 4.4\\
34 4.475\\
35 4.405\\
36 4.245\\
37 3.965\\
38 3.89\\
39 3.825\\
40 3.836\\
41 3.836\\
42 3.818\\
43 3.826\\
44 3.961\\
45 4.096\\
46 4.159\\
47 4.089\\
48 3.974\\
49 3.814\\
50 3.814\\
};
\addlegendentry{2003/03/10};

\addplot [
color=orange,
solid
]
table[row sep=crcr]{
1 5.397\\
2 5.475\\
3 5.515\\
4 5.555\\
5 5.575\\
6 5.55\\
7 5.565\\
8 5.735\\
9 5.904\\
10 6.039\\
11 5.979\\
12 5.769\\
13 5.194\\
14 5.077\\
15 5.092\\
16 5.124\\
17 5.132\\
18 5.097\\
19 5.12\\
20 5.295\\
21 5.47\\
22 5.58\\
23 5.535\\
24 5.345\\
25 4.945\\
26 4.843\\
27 4.838\\
28 4.85\\
29 4.873\\
30 4.863\\
31 4.883\\
32 5.055\\
33 5.228\\
34 5.353\\
35 5.308\\
36 5.138\\
37 4.773\\
38 4.738\\
};
\addlegendentry{2004/03/10};

\addplot [
color=mycolor1,
solid
]
table[row sep=crcr]{
1 6.768\\
2 6.87\\
3 6.945\\
4 7.015\\
5 7.05\\
6 7.061\\
7 7.085\\
8 7.465\\
9 7.835\\
10 8.065\\
11 8.038\\
12 7.865\\
13 6.715\\
14 6.572\\
15 6.592\\
16 6.615\\
17 6.645\\
18 6.62\\
19 6.64\\
20 6.945\\
21 7.225\\
22 7.46\\
23 7.433\\
24 7.245\\
25 6.25\\
26 6.105\\
27 6.13\\
28 6.15\\
29 6.175\\
30 6.15\\
31 6.175\\
32 6.465\\
33 6.74\\
34 6.985\\
35 6.96\\
36 6.765\\
37 5.855\\
38 5.735\\
39 5.755\\
40 5.775\\
41 5.795\\
42 5.78\\
43 5.79\\
44 6.065\\
45 6.33\\
46 6.575\\
47 6.545\\
48 6.34\\
49 5.485\\
50 5.365\\
51 5.395\\
52 5.425\\
53 5.45\\
54 5.43\\
55 5.455\\
56 5.74\\
57 6.015\\
58 6.245\\
59 6.225\\
60 6.035\\
61 5.175\\
62 5.065\\
63 5.09\\
64 5.115\\
65 5.14\\
66 5.14\\
67 5.168\\
};
\addlegendentry{2005/03/10};

\addplot [
color=mycolor2,
solid
]
table[row sep=crcr]{
1 6.646\\
2 6.849\\
3 7.032\\
4 7.215\\
5 7.382\\
6 7.502\\
7 7.657\\
8 8.787\\
9 9.792\\
10 10.482\\
11 10.482\\
12 10.282\\
13 8.712\\
14 8.54\\
15 8.6\\
16 8.675\\
17 8.745\\
18 8.765\\
19 8.835\\
20 9.45\\
21 10.035\\
22 10.465\\
23 10.465\\
24 10.28\\
25 8.34\\
26 8.15\\
27 8.205\\
28 8.26\\
29 8.31\\
30 8.34\\
31 8.41\\
32 9\\
33 9.565\\
34 10.015\\
35 10.015\\
36 9.815\\
37 7.845\\
38 7.645\\
39 7.71\\
40 7.775\\
41 7.83\\
42 7.86\\
43 7.935\\
44 8.52\\
45 9.095\\
46 9.555\\
47 9.56\\
48 9.375\\
49 7.385\\
50 7.185\\
51 7.245\\
52 7.3\\
53 7.35\\
54 7.375\\
55 7.455\\
56 8.05\\
57 8.635\\
58 9.095\\
59 9.09\\
60 8.905\\
61 6.955\\
62 6.785\\
63 6.85\\
64 6.92\\
65 6.985\\
66 6.99\\
67 7.035\\
68 7.575\\
69 8.115\\
70 8.795\\
71 8.79\\
72 8.605\\
};
\addlegendentry{2006/03/10};

\addplot [
color=green,
solid
]
table[row sep=crcr]{
1 6.912\\
2 7.055\\
3 7.19\\
4 7.328\\
5 7.435\\
6 7.492\\
7 7.6\\
8 8.225\\
9 8.84\\
10 9.15\\
11 9.145\\
12 8.915\\
13 7.64\\
14 7.517\\
15 7.57\\
16 7.635\\
17 7.695\\
18 7.747\\
19 7.842\\
20 8.312\\
21 8.767\\
22 9.012\\
23 9.007\\
24 8.787\\
25 7.417\\
26 7.297\\
27 7.362\\
28 7.432\\
29 7.487\\
30 7.537\\
31 7.637\\
32 8.067\\
33 8.482\\
34 8.712\\
35 8.717\\
36 8.492\\
37 7.232\\
38 7.112\\
39 7.167\\
40 7.237\\
41 7.282\\
42 7.332\\
43 7.437\\
44 7.862\\
45 8.282\\
46 8.507\\
47 8.512\\
48 8.297\\
49 7.047\\
50 6.932\\
51 6.997\\
52 7.072\\
53 7.132\\
54 7.182\\
55 7.282\\
56 7.692\\
57 8.097\\
58 8.317\\
59 8.322\\
60 8.097\\
61 6.897\\
62 6.787\\
63 6.852\\
64 6.917\\
65 6.967\\
66 7.017\\
67 7.107\\
68 7.497\\
69 7.887\\
70 8.147\\
71 8.152\\
72 7.927\\
};
\addlegendentry{2007/03/12};

\addplot [
color=mycolor3,
solid
]
table[row sep=crcr]{
1 10.024\\
2 10.075\\
3 10.136\\
4 10.214\\
5 10.27\\
6 10.268\\
7 10.32\\
8 10.535\\
9 10.84\\
10 11.045\\
11 10.99\\
12 10.71\\
13 8.99\\
14 8.84\\
15 8.882\\
16 8.934\\
17 8.987\\
18 8.995\\
19 9.075\\
20 9.325\\
21 9.69\\
22 9.915\\
23 9.91\\
24 9.67\\
25 8.54\\
26 8.45\\
27 8.5\\
28 8.56\\
29 8.61\\
30 8.615\\
31 8.695\\
32 8.97\\
33 9.335\\
34 9.565\\
35 9.559\\
36 9.319\\
37 8.419\\
38 8.379\\
39 8.439\\
40 8.504\\
41 8.554\\
42 8.564\\
43 8.644\\
44 8.924\\
45 9.294\\
46 9.524\\
47 9.524\\
48 9.299\\
49 8.439\\
50 8.414\\
51 8.474\\
52 8.534\\
53 8.584\\
54 8.599\\
55 8.669\\
56 8.949\\
57 9.314\\
58 9.544\\
59 9.549\\
60 9.319\\
61 8.509\\
62 8.494\\
63 8.554\\
64 8.614\\
65 8.664\\
66 8.674\\
67 8.739\\
68 9.019\\
69 9.379\\
70 9.614\\
71 9.619\\
72 9.394\\
};
\addlegendentry{2008/03/10};

\addplot [
color=mycolor4,
solid
]
table[row sep=crcr]{
1 3.84\\
2 3.907\\
3 4.027\\
4 4.176\\
5 4.27\\
6 4.318\\
7 4.423\\
8 4.933\\
9 5.461\\
10 5.736\\
11 5.751\\
12 5.636\\
13 5.361\\
14 5.389\\
15 5.479\\
16 5.594\\
17 5.668\\
18 5.697\\
19 5.791\\
20 6.111\\
21 6.486\\
22 6.716\\
23 6.716\\
24 6.526\\
25 6.006\\
26 5.981\\
27 6.056\\
28 6.146\\
29 6.216\\
30 6.246\\
31 6.331\\
32 6.576\\
33 6.856\\
34 7.076\\
35 7.076\\
36 6.841\\
37 6.186\\
38 6.131\\
39 6.196\\
40 6.291\\
41 6.361\\
42 6.386\\
43 6.466\\
44 6.701\\
45 6.991\\
46 7.211\\
47 7.211\\
48 6.971\\
49 6.306\\
50 6.261\\
51 6.331\\
52 6.416\\
53 6.491\\
54 6.521\\
55 6.601\\
56 6.836\\
57 7.121\\
58 7.346\\
59 7.346\\
60 7.106\\
61 6.366\\
62 6.316\\
63 6.391\\
64 6.481\\
65 6.546\\
66 6.576\\
67 6.666\\
68 6.896\\
69 7.176\\
70 7.396\\
71 7.391\\
72 7.156\\
};
\addlegendentry{2009/03/10};

\addplot [
color=mycolor5,
solid
]
table[row sep=crcr]{
1 4.559\\
2 4.624\\
3 4.696\\
4 4.789\\
5 4.851\\
6 4.886\\
7 4.981\\
8 5.286\\
9 5.641\\
10 5.851\\
11 5.811\\
12 5.681\\
13 5.353\\
14 5.345\\
15 5.388\\
16 5.444\\
17 5.5\\
18 5.533\\
19 5.635\\
20 5.887\\
21 6.177\\
22 6.385\\
23 6.345\\
24 6.15\\
25 5.645\\
26 5.619\\
27 5.664\\
28 5.719\\
29 5.769\\
30 5.804\\
31 5.909\\
32 6.154\\
33 6.434\\
34 6.639\\
35 6.604\\
36 6.384\\
37 5.824\\
38 5.779\\
39 5.829\\
40 5.899\\
41 5.954\\
42 5.989\\
43 6.099\\
44 6.352\\
45 6.63\\
46 6.835\\
47 6.81\\
48 6.59\\
49 6.02\\
50 5.98\\
51 6.035\\
52 6.11\\
53 6.17\\
54 6.205\\
55 6.315\\
56 6.565\\
57 6.845\\
58 7.05\\
59 7.025\\
60 6.805\\
61 6.225\\
62 6.18\\
63 6.235\\
64 6.31\\
65 6.375\\
66 6.41\\
67 6.515\\
68 6.765\\
69 7.045\\
70 7.25\\
71 7.23\\
72 7.015\\
};
\addlegendentry{2010/03/10};

\addplot [
color=blue,
solid
]
table[row sep=crcr]{
1 3.83\\
2 3.892\\
3 3.955\\
4 4.03\\
5 4.078\\
6 4.098\\
7 4.147\\
8 4.335\\
9 4.593\\
10 4.726\\
11 4.717\\
12 4.668\\
13 4.548\\
14 4.581\\
15 4.616\\
16 4.658\\
17 4.688\\
18 4.698\\
19 4.744\\
20 4.914\\
21 5.141\\
22 5.269\\
23 5.244\\
24 5.159\\
25 4.926\\
26 4.943\\
27 4.973\\
28 5.015\\
29 5.045\\
30 5.06\\
31 5.117\\
32 5.262\\
33 5.484\\
34 5.614\\
35 5.589\\
36 5.496\\
37 5.256\\
38 5.261\\
39 5.286\\
40 5.331\\
41 5.366\\
42 5.376\\
43 5.436\\
44 5.583\\
45 5.808\\
46 5.945\\
47 5.927\\
48 5.837\\
49 5.592\\
50 5.594\\
51 5.619\\
52 5.664\\
53 5.707\\
54 5.717\\
55 5.777\\
56 5.914\\
57 6.134\\
58 6.267\\
59 6.247\\
60 6.137\\
61 5.882\\
62 5.877\\
63 5.897\\
64 5.937\\
65 5.972\\
66 5.992\\
67 6.064\\
68 6.204\\
69 6.419\\
70 6.554\\
71 6.531\\
72 6.421\\
};
\addlegendentry{2011/03/10};

\addplot [
color=mycolor6,
solid
]
table[row sep=crcr]{
1 2.299\\
2 2.408\\
3 2.507\\
4 2.602\\
5 2.655\\
6 2.677\\
7 2.74\\
8 2.94\\
9 3.267\\
10 3.404\\
11 3.414\\
12 3.397\\
13 3.368\\
14 3.407\\
15 3.462\\
16 3.506\\
17 3.52\\
18 3.52\\
19 3.557\\
20 3.661\\
21 3.872\\
22 3.98\\
23 3.97\\
24 3.92\\
25 3.8\\
26 3.816\\
27 3.843\\
28 3.88\\
29 3.896\\
30 3.899\\
31 3.935\\
32 4.024\\
33 4.216\\
34 4.315\\
35 4.292\\
36 4.223\\
37 4.068\\
38 4.081\\
39 4.107\\
40 4.142\\
41 4.158\\
42 4.161\\
43 4.196\\
44 4.283\\
45 4.475\\
46 4.577\\
47 4.551\\
48 4.481\\
49 4.311\\
50 4.323\\
51 4.349\\
52 4.386\\
53 4.406\\
54 4.41\\
55 4.444\\
56 4.531\\
57 4.724\\
58 4.821\\
59 4.796\\
60 4.721\\
61 4.561\\
62 4.571\\
63 4.596\\
64 4.631\\
65 4.655\\
66 4.661\\
67 4.698\\
68 4.793\\
69 4.988\\
70 5.088\\
71 5.061\\
72 4.986\\
};
\addlegendentry{2012/03/13};

\addplot [
color=mycolor7,
solid
]
table[row sep=crcr]{
1 3.649\\
2 3.689\\
3 3.734\\
4 3.787\\
5 3.808\\
6 3.81\\
7 3.837\\
8 3.928\\
9 4.103\\
10 4.192\\
11 4.183\\
12 4.129\\
13 4.03\\
14 4.044\\
15 4.066\\
16 4.096\\
17 4.113\\
18 4.118\\
19 4.152\\
20 4.229\\
21 4.394\\
22 4.482\\
23 4.468\\
24 4.379\\
25 4.169\\
26 4.177\\
27 4.196\\
28 4.229\\
29 4.247\\
30 4.251\\
31 4.285\\
32 4.355\\
33 4.525\\
34 4.625\\
35 4.605\\
36 4.521\\
37 4.296\\
38 4.31\\
39 4.333\\
40 4.37\\
41 4.388\\
42 4.391\\
43 4.426\\
44 4.514\\
45 4.701\\
46 4.801\\
47 4.781\\
48 4.699\\
49 4.464\\
50 4.478\\
51 4.501\\
52 4.538\\
53 4.556\\
54 4.559\\
55 4.594\\
56 4.699\\
57 4.899\\
58 5.004\\
59 4.984\\
60 4.904\\
61 4.681\\
62 4.698\\
63 4.725\\
64 4.763\\
65 4.782\\
66 4.787\\
67 4.824\\
68 4.942\\
69 5.158\\
70 5.263\\
71 5.243\\
72 5.163\\
};
\addlegendentry{2013/03/11};

\end{axis}
\end{tikzpicture}%
\caption{Futures curve for Nymex NG at different observation dates} 
\label{fig:FutureCurveNG} 
\end{figure}

As mentioned above, the first main feature of natural gas prices is constituted by  the presence of
 a seasonality component: we plot the NG futures curve for several dates in Figure \ref{fig:FutureCurveNG}. As noted above, we remark repeated rises
 in prices that occur during every winter, which are clearly due to the 
demand for heating during cold periods of the year. In addition to this traditional seasonal feature, the use of natural gas for electricity generation has created a second smaller increase during the summer period, because of the cooling systems during warm periods. These expected patterns in natural gas prices are the first source of value of a storage unit, since it is possible to buy summer futures contracts and store the gas delivered during summer, and sell more valuable winter futures contracts, which implies gas withdrawal during cold periods. This is the basic idea behind the so-called "intrinsic strategy," which is based only on futures contracts and exploits the calendar spreads in the futures curve (see Appendix \ref{appB} for more details about the intrinsic value of a storage unit.)\\

\begin{figure} 
\setlength\figureheight{8cm} 
\setlength\figurewidth{15cm}
\input{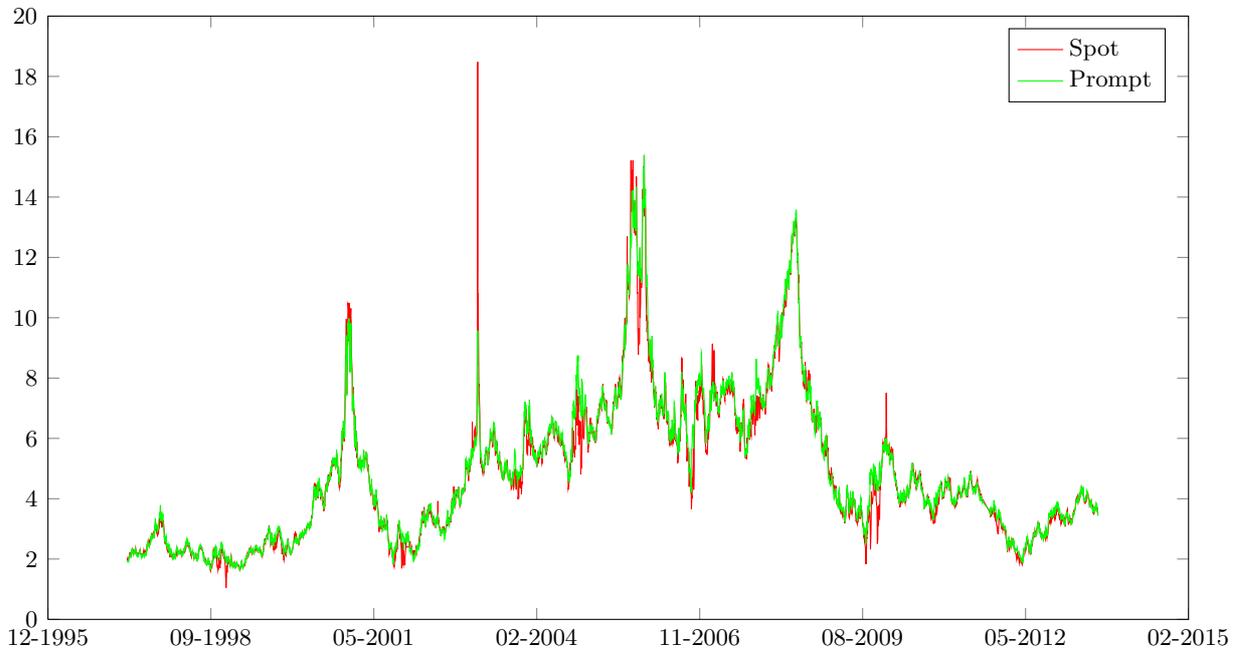} 
\caption{Spot and prompt historical prices} 
\label{fig:SpotPromptHistorical} 
\end{figure}



The second important aspect of natural gas prices, is the presence of sudden moves due to unexpected rise in demand (caused in general by an unpredicted weather change), technical problems in the supply chain, or a poor anticipation of 
the global stored gas available in the market. Because of delivery constraints (one day) of spot contracts, such sudden changes are almost instantaneously
 reflected in the spot dynamics, giving  rise to large shifts in prices, 
rapidly absorbed by the storage capacities available in the market. 
These large and quickly absorbed jumps, commonly called spikes, can be viewed in Figure \ref{fig:SpotPromptHistorical}, which shows
 many sudden dislocations between spot and prompt prices.
 \footnote{We use a 1997-2013 historical data of spot and prompt price, published by the U.S. Energy Information Administration. cf \url{http://www.eia.gov/dnav/ng/ng_pri_fut_s1_d.htm}}
For example, we can notice a large spike in the spot price 
during late February 2003, when the natural gas price jumped by almost $78.00\%$ and $54.26\%$  in two successive days, then went down by $-43.34\%$ and $-19.58\%$ during the two following days. 
As noted by the US Federal Energy Regulatory Commission (FERC) by \cite{FERCSpike2003}, this spike in gas price was due to \enquote{physical market conditions leading to low supply and high demand for a short time.}
\cite{FERCSpike2003} also observed that
\enquote{similar natural gas price spikes are possible when episodes of cold weather occur at times when 
 storage inventories are limited.}\\

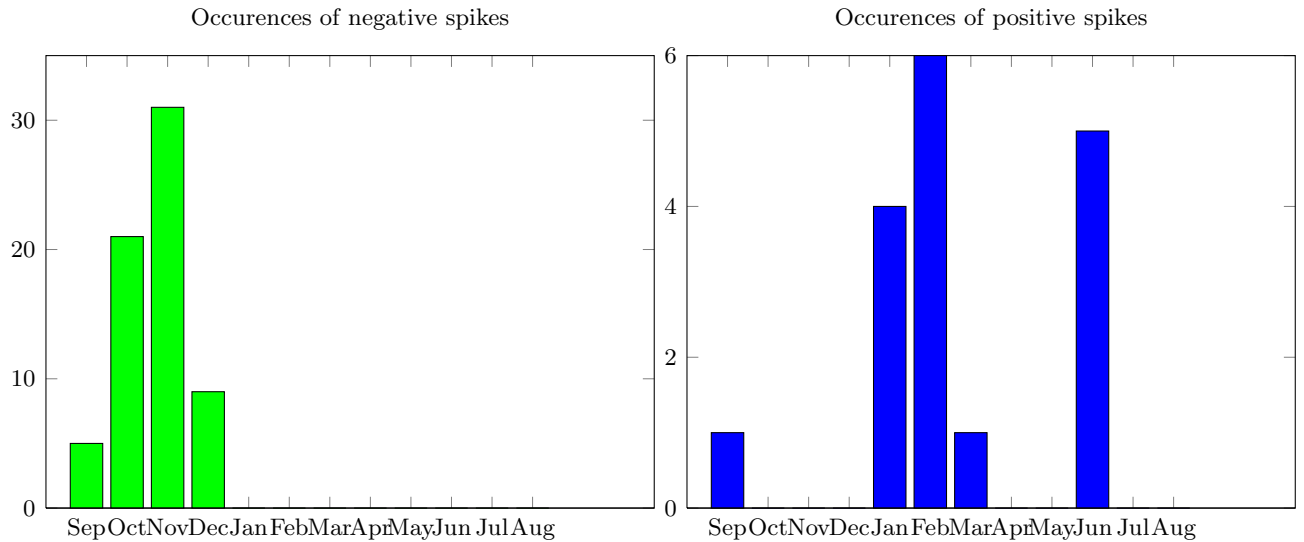
\begin{figure} 
\setlength\figureheight{6cm} 
\setlength\figurewidth{8cm}
%
%
%
%
\begin{tikzpicture}

\begin{axis}[%
width=\figurewidth,
height=\figureheight,
area legend,
scale only axis,
xmin=0,
xmax=15,
xtick={1,2,3,4,5,6,7,8,9,10,11,12},
xticklabels={Sep,Oct,Nov,Dec,Jan,Feb,Mar,Apr,May,Jun,Jul,Aug},
ymin=0,
ymax=35,
name=plot1,
title={Occurences of negative spikes}
]
\addplot[ybar,bar width=0.0533333333333333\figurewidth,fill=green,draw=black] plot coordinates{(1,5)
(2,21)
(3,31)
(4,9)
(5,0)
(6,0)
(7,0)
(8,0)
(9,0)
(10,0)
(11,0)
(12,0)};

\addplot [
color=black,
solid,
forget plot
]
table[row sep=crcr]{
0 0\\
15 0\\
};
\end{axis}

\begin{axis}[%
width=\figurewidth,
height=\figureheight,
area legend,
scale only axis,
xmin=0,
xmax=15,
xtick={1,2,3,4,5,6,7,8,9,10,11,12},
xticklabels={Sep,Oct,Nov,Dec,Jan,Feb,Mar,Apr,May,Jun,Jul,Aug},
ymin=0,
ymax=6,
at=(plot1.right of south east),
anchor=left of south west,
title={Occurences of positive spikes}
]
\addplot[ybar,bar width=0.0533333333333333\figurewidth,fill=blue,draw=black] plot coordinates{(1,1)
(2,0)
(3,0)
(4,0)
(5,4)
(6,6)
(7,1)
(8,0)
(9,0)
(10,5)
(11,0)
(12,0)};

\addplot [
color=black,
solid,
forget plot
]
table[row sep=crcr]{
0 0\\
15 0\\
};
\end{axis}
\end{tikzpicture}%
\caption{Occurrences of spikes} 
\label{fig:SpikesOccurrence} 
\end{figure}

The appearance of the  spikes is correlated to the spot and prompt prices
 spread. Indeed, while the prompt contract is a good proxy for the spot price, it does not suffer from sudden shocks of the same amplitude as spot prices, because of time-to-maturity factor. 
The spikes provoke 'unusual' gaps between the two contracts. 
In our study, we detect spikes by identifying the outliers from the time series $(x_t)$  of the spread between the spot price $S_t$ and the prompt price $P_t$ given by 
$x_t:=\frac{S_t-P_t}{P_t}$;  we separate the study of positive and negative
 spikes, since they reflect two different market conditions. In fact, 
positive spikes are often caused by unpredicted weather changes, such as
 a cold front or a heat wave. On the other hand, negative spikes are generally due to a poor anticipation of market-wide gas storage levels. In Figure \ref{fig:SpikesOccurrence} we plot the number of occurrences of negative and positive spikes during each month.
We remark that the repartition of the spikes is clearly dependent on their sign. In fact, most of the positive spikes happen during the winter months of January and February and the summer month of June, which can be explained by the
 occurrence of an unpredicted cold front or heat wave. On the other hand, the negative spikes appear during the months of pre-heating periods, i.e. September, October and November. One plausible explanation is given by \cite{eia2007}, which states:
 \enquote{October is the last month of the refill season. There may be increased competition from storage facilities looking to meet end-of-season refill goals as well as increased anticipation regarding the upcoming heating season.}\\

In order to take into account the stylized facts mentioned above, first, 
our futures model  incorporates seasonality in the futures curve,
and second, the spot model  describes the existence of spikes 
and takes into account  the correlation between
 spot and futures prices, through the prompt contract.
 As far as our  knowledge is concerned, these two facts were not taken
 into account in the literature related to gas storage valuation; 
 we believe that they constitute the main sources of storage value.
 Indeed, storage managers can exploit the  winter/summer seasonality  of gas prices,  buying during  summer and selling  during winter.
 The presence of spikes can be monetized (if  the injection/withdrawal
 rates are high enough), which is possible for some high deliverability storage units (see Appendix A.)

Before discussing modeling issues, we start by specifying the gas storage unit valuation and hedging problem,  and recall numerical simulation algorithms.

\section{Valuation and hedging of a gas storage utility}
\label{S3}
The problem of valuing gas storage units has been discussed from many angles in the literature, yielding different approaches and numerical methods. Leasing a gas storage unit is equivalent to paying for the right, but not the obligation, to inject or withdraw gas from the unit. Hence the goal of the owner is to optimize the use of the gas storage facility, by injecting or withdrawing gas from the unit and, at the same time, trading gas on the spot and/or futures market. All these decisions have to be made under many operational constraints, such as
 maximal and minimal volume of the storage, and limited injection and withdrawal rates. This forces the resolution of a constrained stochastic control problem.\\

The gas spot price is modeled by a process, denoted by $S$. We suppose that this process is given as a function of a Markov process $X$ in term of which the optimal control problem will be expressed. For example, in the framework
 \eqref{dejongDyn}, used by \cite{boogert2008gas}, the spot process is a Markov process, so we take obviously $X = S$. 

\cite{warin2010gas} takes the futures curve as the underlying process, such that
\begin{eqnarray} \label{EFuture}
F(t,T) = F(0,T) \exp \left[ -\frac{1}{2} V(t,T) + \Sum_{i=1}^n  e^{-a_i(T-t)} W^i_t \right]  \; , 
\end{eqnarray} with $ W^i_t = \int_0^t \sigma_i(u) e^{-a_i(T-u)} dZ^i_u$, 
$Z^i$ being  standard Wiener processes, and $
V(t,T) = var(\Sum_{i=1}^n  e^{-a_i(T-t)} W^i_t )$.
Indeed, the Markov process $X$ can be chosen to be equal to the random sources 
 $X = (W^1, ..., W^n)$.

In fact the futures prices modeled in \eqref{EFuture} are martingales.
In particular one supposes that the underlying probability is risk-neutral
 probability, and not necessarily unique.

\begin{remark}
For simplicity, in the rest of this article, we suppose the discount interest rate to vanish, and consider the problem specification in a time discrete setting.\\
\end{remark}

In the next part of the paper we present the specification of the gas storage valuation problem, using the notations of \cite{warin2010gas}.

\subsection{Gas storage specification}
\label{S31}

We consider a gas storage facility with technical constraints (either physical or regulatory) on the volume of stored gas, $V_{min}$ and $V_{max}$ i.e. at each time, the volume level of the stored gas $V$ should verify $V_{min} \leq V \leq  V_{max}$.\\

We suppose a discrete set of dates $t_i=i \Delta t$ for $i=0,...,n-1$ with $\Delta t=T/n$. At each date $t_i$ and starting from a volume $V_{t_i}$, the user has the possibility to make one of three decisions: either inject gas at rate of $a_{inj}$, or withdraw gas at rate of $a_{with}$ or take no action. We denote by $u_i$ the decision at time $t_i$, and write $u_i=inj$ (resp. $with$, $no$) if the decision is injecting gas at rate $a_{inj}$ (resp. withdrawing gas at rate $a_{with}$, no action). \\

If the user decides to follow a strategy $(u_i)_{i=0...n-1}$, then the volumes of gas in the storage $(V_i)_i$ are given by the iteration
\begin{eqnarray}
V_0 &=& v, \\
V_{t_{i+1}}(u) &=& \left\{ \begin{array}{ll}
 \min(V_{t_i}(u) + a_{inj} \Delta t, V_{max}) &\mbox{ if $u_i=ing$} \\
 \max(V_{t_i}(u) - a_{with} \Delta t, V_{min}) &\mbox{ if $u_i=with$} \\
 V_{t_i}(u) &\mbox{ if $u_i=no$},
       \end{array} \right.
\label{volume}
\end{eqnarray}
for $i=0,...,n-2$. The generated cash flow -selling, in the case of gas withdrawal, or buying, in the case of gas injection- is given by
$$
\phi_{u_i}(S_{t_i}) := S_{t_i} (V_{t_{i+1}}(u)-V_{t_i}(u)).
$$
In general the maximum and minimum injection and withdrawal speeds ($a_{inj}$ and $a_{with}$) are functions of the amount of gas in storage. However, without loss of generality, we assume for simplicity that these bounds are constant. 
In Table \ref{TableA} we summarize the possible decisions and their consequences on gas volume and generated cash flow.
\begin{table}
\begin{tabular}{|p{4cm}|p{7cm}|p{4.8cm}|} 
\hline
\textbf{Decision $u$} & \textbf{Next volume} & \textbf{Cash flow} \\
\hline
Injection: $u_i=inj$ & $ V_{t_{i+1}}(u) = \min(V_{max}, V_{t_i}(u) + a_{inj} \Delta t)$ & $\phi_{inj}=S_{t_i} (V_{t_i}(u)-V_{t_{i+1}}(u))$\\
\hline
Withdrawal: $u_i=with$ & $V_{t_{i+1}}(u) = \max(V_{min}, V_{t_i}(u) - a_{with} \Delta t)$ & $\phi_{with}=S_{t_i} (V_{t_i}(u)-V_{t_{i+1}}(u))$\\
\hline
No Action: $u_i=no$ & $ V_{t_{i+1}}(u) = V_{t_i}(u) $ & $\phi_{no}=0$\\
\hline
\end{tabular}
\caption{Possible decisions}
\label{TableA}
\end{table}
We recall that $a_{inj}$ indicates the injection rate per time unit $\Delta t$
and  $a_{with}$ the withdrawal rate per time unit $\Delta t$.
Consequently, the wealth generated by following a strategy $u$ is given by
\begin{eqnarray}
\text{Wealth}_{\text{spot}}(u) = \Sum_{i=0}^{n-1} \phi_{u_i}(S_{t_i}).
\label{cumWealth}
\end{eqnarray}

Finally, we are interested in the expectation of this cumulative cash flows, which we denote by $J$.
More precisely we set
\begin{eqnarray}
J(t_0, x_0, v_0 ; u) &:=& \E\big[ \text{Wealth}_{\text{spot}}\big]\\
 &=& \E\big[ \Sum_{i=0}^{n-1} \phi_{u_i}(S_{t_i})\big] \nonumber ,
\end{eqnarray}
$J$ is a function that depends on the initial time $t_0$, the value of the Markov process $X_0=x_0$, the initial volume in the storage 
$v_0$  and the strategy $u$.\\

The goal of the storage operator is to find a strategy $u$ maximizing the expected cumulative cash flows. We denote this optimal value by $J^{\star}$. So the problem to solve is the following:
\begin{eqnarray} 
J^{\star}(t_0, x_0, v_0) &=& \max_{(u_i)_{i=0...n-1}} J(t_0, x_0, v_0 ; u) = 
\max_{(u_i)_{i=0...n-1}} \E\big[ \Sum_{i=0}^{n-1} \phi_{u_i}(S_{t_i})\big] \nonumber \\
&& \\
&=& J(t_0, x_0, v_0; u^{\star}). \nonumber
\label{maxPb}
\end{eqnarray}
A priori, the underlying probability is the historical probability measure,
 at least if the futures do not intervene in the spot model.
This quantity constitutes an objective for the manager. However, it is
not a ``fair'' price in the sense of ``absence of arbitrage,''
 since the spot is not traded as a  financial asset; 
in that case the price would be an expectation with respect 
 to a  risk-neutral probability.
  On the other hand, $J^{\star}$ constitutes a price indicator;
 practitioners 
trade gas storage units at a proportion of this price. \\

In our proposed framework (see Section \ref{S5}) the spot and the
futures are jointly modeled on a product  space
$\Omega = (\Omega_s,\Omega_f)$ equipped with a probability $\Q$.
 The futures are first directly described
as martingales on $\Omega_f$  with respect to their corresponding
 risk-neutral probability $\P^\ast$ and they are extended trivially
to $\Omega$. Formally $\Q$ is defined by
$ \Q(d\omega_s,d\omega_f) = Q^{\omega_f} (d\omega_s) P^\ast(d\omega_f)$,
where $Q^{\omega_f}(d\omega_s)$ is a random probability kernel
(historically considered), describing the random behavior
of $S$ for each realization $\omega_f$ of the futures asset. 
The expectation of the optimal cumulative cash flows 
with respect to $Q$ will be
then  a price indicator, compatible
with classic financial principles, as far as  
futures assets are concerned. 
Indeed, we will also estimate the volatility parameters for
the diffusion describing the futures assets
 $F$ using historical data, that is 
under some historical probability $\P$ and not $\P^\ast$
as we would need. However, 
the probability $\P^\ast$ is equivalent to 
 $\P$ and this justifies the coherence of the estimation.


\subsection{Dynamic programming equation}

\label{SDPE}

From Table \ref{TableA}, we recall that $ V_{t_{i+1}}(u)$ only depends on $V_{t_i}(u)$ and $u_i$.
To emphasize this fact, if $V_{t_i}=v$, we also express  $ V_{t_{i+1}}(u)$ by $\widehat{V}_{u_i}(v)$. 

At time $t$, for $X_t=x$  and with current volume level $v$, the (optimal) value for gas storage
will be of course denoted by $J^{\star}(t, x, v)$.
The dynamic programming principle implies 
\begin{eqnarray}
J^{\star}(t_i, x, v) = \max_{u_i \in\{inj,~no,~with\}} \big\{\phi_{u_i} + \E\big[J^{\star}(t_{i+1}, X_{t_{i+1}}, \widehat{V}_{u_i}(v)) | X_{t_i}=x, V_{t_i}=v\big] \big\}.
\label{dyn}
\end{eqnarray}

The classic way to solve this problem numerically is to use Monte Carlo simulations, combined with the \cite{LongstaffSchwartz} algorithm,
 which approximates the above conditional expectation, using a regression technique. This backward algorithm yields an estimate of the optimal strategy. As noted by \cite{boogert2008gas}, the main difficulty comes from the fact that the value function
also depends on volume level, which in turn depends on the optimal strategy.\\

To circumvent this difficulty, \cite{boogert2008gas} suggest discretizing the volume into a finite grid, $v_l=V_{min}+l\delta, \quad l=0,...,L= (V_{max}-V_{min})/\delta$ where $L$ is the number of volume subintervals. However, the fact that the
 time grid is discrete and that at each time the storage unit manager has only three possible actions implies that the number of attainable volumes for any strategy is finite. In fact, at each time $t_i$, the set $\mathcal{V}(i)$ of possible volumes is given by  
$$\mathcal{V}(i)=\{ V_i = v_0 + k a_{inj} \Delta t + l a_{with} \Delta t \text{ , such that }
 V_{min} \leq V_i \leq V_{max} \text{ and } k,l\in \N \;, k+l\leq i\};$$ 
consequently, it is possible to solve the dynamic programming equation \eqref{dyn} for all volumes in $\mathcal{V}(i)$. 
The only motivation to use a restricted volume grid would be the reduction of computation time.\\

We then get the following equation at time $t_i$, for each path simulation $X^m$, where $m = 1,\ldots, M$,
$M$ being the total number of realization paths,
 and each volume level $v_l\in \mathcal{V}(i)$:
\begin{eqnarray}
J^{\star}(t_i, X^{m}_{t_i}, v_l) = \max_{u_i \in\{inj,~no,~with\}} \big\{\phi_{u_i} + \E\big[J^{\star}(t_{i+1}, X_{t_{i+1}}, \widehat{V}_{u_i}(v_l)) | X_{t_i}=X^{m}_{t_i}, V_{t_i}=v_l\big] \big\}.
\label{dyn_vol}
\end{eqnarray}

The conditional expectation above is estimated using Longstaff-Schwartz regression algorithm, at each volume grid point $v_l$.\\

This will give us an estimation of the optimal strategy, denoted $u^{\star}$, and the initial value of the gas storage unit
 $J^{\star}(t_0, X_{t_0}, v_0) = J(t_0, X_{t_0}, v_0; u^{\star})$. The numerical resolution of  problem \eqref{dyn} is done in two phases.
 The first stage consists of estimating the optimal strategy $u^{\star}$, by performing the backward iterations of equation \eqref{dyn_vol}, using regression techniques to estimate the conditional expectations, along a set of simulated paths. 
The second phase consists of estimating the value function $J^{\star}$ through the forward iterations of \eqref{dyn_vol},
 along a new set of simulated paths, where we apply the estimated optimal strategy, as given by the backward algorithm. \\

One important remark about  problem \eqref{maxPb} is that the maximization is carried out for the expected wealth 
generated by the spot-trading strategy. Consequently, following the corresponding  estimated optimal strategy
on a single path will not ensure that the manager will recover the initial storage value $J^{\star}$. 
  There will certainly be a discrepancy between the  realized cumulative cash flows on a given  path and  the expected value $J^{\star}$. 
Hence, it is crucial for the storage manager to reduce the variance of the cumulative cash flows, which is a random variable.
 As we will explain in the next section, this will be realized by conducting a financial hedging strategy, based on futures contracts on natural gas. 

\subsection{Financial hedging strategy}
\label{hedging}

After estimating  the optimal gas strategy, the storage unit manager will follow these optimal decisions on the sample path revealed by the market. But one should keep in mind that, if one follows this optimal strategy $u^{\star}$, the cumulative wealth 
is only the realization of a random variable whose expectation equals
 the initial price $J^{\star}$ of the gas storage unit.  This motivates the interest in hedging strategies enabling better tracking of the storage value. This can be done by combining optimal gas strategies and 
 additional financial trades, so that the expectation of the related  cumulative  wealth generated by both physical and financial operations 
is still $J^{\star}$, but its variance (or some other risk criterion) is reduced. Analogously to
  \cite{bjerksund2011}, who treats the  intrinsic value case,
this additional financial hedging strategy plays a similar role to 
the control variate in the variance reduction of Monte Carlo 
simulations, as it preserves the expected cumulative cash flows and reduces its variance. 
In general to reduce the variance of a Monte Carlo estimator of a r.v. $Y$, one adds to it a mean zero control variate, which is
  highly (negatively) correlated to $Y$. Since futures contracts are the most liquid assets in the natural gas market, and are strongly correlated to the spot price, 
they form an ideal hedging instrument. In fact, although a futures contract price $F(t,T)$ does not  converge  to the spot
 price, when the time to maturity $T-t$ goes to zero,  the correlation between the prompt contract (for example) and the spot price is very high, and often the two contracts move in the same direction.  In practice, in the market, one has access to a set of  futures standard contracts, 
with specified maturities $\{T_j\}_{1 \leq j \leq m}$. The basic idea of a financial hedging strategy is to add to the physical spot 
trading, a strategy of buying and selling, at a trading date $t_i$, a quantity $\Delta(t_i, T_j)$ of futures contracts $F(.,T_j)$ 
for $1 \leq j \leq m$. Logically, those quantities will depend on the spot and futures prices, 
$S$ and $\{F(.,T_j)\}_{1 \leq j \leq m}$, but also on the current volume level.\\ 

If the gas storage manager follows such a hedging strategy, in addition to the spot physical trading, then the cumulative cash
 flows of those two combined strategies is equal to
\begin{eqnarray}\label{cashFlowHedg}
\text{Wealth}_{\text{spot+futures}} = \Sum_{i=0}^{n-1} \phi_{u^{\star}_i}(S_{t_i}) + \Sum_{i=0}^{n-1}  \Sum_{j=1}^{m}  \Delta(t_i, T_j) ( F(t_{i+1}, T_j) - F(t_i, T_j)).
\end{eqnarray}
Because the futures contract $F(., T_j)$ stops trading after its expiration date $T_j$, we use the convention 
$\Delta(t, T_j)=0 \text{, for } t \geq T_j$.\\

Since the futures price process is a martingale under the risk neutral probability, we have $\E_{t_i}\big[ F(t_{i+1}, T_j) \big]= F(t_i, T_j)$. Hence, the expectation of this hedging strategy is null i.e. $$\E\big[\Sum_{i=0}^{n-1}  \Sum_{j=1}^{m}  \Delta(t_i, T_j) ( F(t_{i+1}, T_j) - F(t_i, T_j))\big] =0 .$$ Consequently, following the optimal spot strategy in parallel with a futures hedging portfolio gives the same cash flows in expectation, but very likely with lower variance.
$$
\E\big[ \text{Wealth}_{\text{spot+futures}} \big] = \E\big[ \text{Wealth}_{\text{spot}} \big] = \E\big[ \Sum_{i=0}^{n-1} \phi_{u^{\star}_i}(S_{t_i}) \big] = J^{\star}.
$$
The specification of such a hedging strategy will of course depend on the nature of the relation between the spot price and 
the futures curve. \\

A heuristic strategy that is widely used in the industry is to take the quantity $\Delta_1(t_i, T_j)$ of futures $F(., T_j)$ to be equal to the conditional expectation of volume to be exercised during the delivery period of the futures contract, conditional on the information at $t_i$. More precisely, the heuristic delta is equal to the $t_i$-conditional expectation
 \begin{equation}\label{EDelta1}
\Delta_1(t_i, T_j) = \E_{t_i}[ \Sum_{T_{j-1} \leq t_l <T_j} V_{l+1}(u^{\star})-V_{l}(u^{\star})].
\end{equation}
We also propose a modification of this heuristic delta, where we use the concept of tangent process (\cite{warin2010gas}).
If we assume that the prompt converges towards the spot, then we can write
\begin{eqnarray*}
\text{Wealth}_{\text{spot}} &=& \Sum_{i=0}^{n-1} \left(  V_{i+1}(u^{\star})-V_{i}(u^{\star}) \right)S_{t_i} \\
&\simeq & \Sum_{i=0}^{n-1} \left(  V_{i+1}(u^{\star})-V_{i}(u^{\star}) \right) P_{t_i} \\
&=& \Sum_{j} \Sum_{T_{j-1} \leq t_l <T_j} \left(  V_{l+1}(u^{\star})-V_{l}(u^{\star}) \right) F(t_l, T_j).
\end{eqnarray*}
So, another heuristic delta $\Delta_2$ can be defined, using the concept of tangent process
\begin{equation}\label{EDelta2}
\Delta_2(t_i, T_j) = \E_{t_i}[ \Sum_{T_{j-1} \leq t_l <T_j} \left(  V_{l+1}(u^{\star})-V_{l}(u^{\star}) \right) \frac{F(t_l, T_j)}{F(t_i, T_j)}]
\end{equation}  We emphasize that the definition of these two hedging strategies is based on heuristic reasoning. Therefore, the hedging will not be perfect and a residual risk will still remain.\\

As we will see in the numerical experiments, this financial hedging strategy allows for a significant reduction of the cash flows uncertainty of the spot trading strategy. In fact, the variance of the spot trading is quite reduced while conducting a futures hedging strategy, and an out-of-sample test applied over a price history of 10 years shows better wealth tracking for the hedging strategy.\\

In the next section we will present several approaches for spot and futures prices modeling, and study the consequences of a model choice.\\

\section{Literature on price processes}
\label{S4}

Generally, the problem of gas storage unit valuation has been studied from the angle of numerical resolution, and not much interest has been paid to the modeling itself and its effects on the final output of the numerical scheme. In fact, we encountered two modeling approaches in the literature. \\

The first approach consists of modeling only the spot model, with classical
 mean-reverting models, as proposed by \cite{boogert2008gas}, with no spike features. This approach, while realistic, does not take into account the dependence of the problem with respect to the futures curve and its dynamics, and does not offer the possibility of a hedging strategy based on futures. 
The second approach is based on the modeling of the futures curve by multi-factor log-normal dynamics, and makes the assumption of considering the spot price equal to the limit of futures prices with time to maturity going to zero. We note, however, that this assumption does not conform to the commodities market. On the other hand, it enables the definition of a delta hedging strategy based on available futures contracts on the market.
We give more details about these two approaches in what follows.\\

A very common  framework consists of modeling
 the spot price as a mean-reverting process. For instance, \cite{boogert2008gas} developed a Monte Carlo method for storage valuation, using the Least Square Monte Carlo method, as proposed by \cite{LongstaffSchwartz} for American options. 
They consider one factor model for the spot price, 
 which is calibrated to the initial futures curve. The price process $S$ is
 given by
\begin{eqnarray}
\frac{dS_t}{S_t} = \kappa [\mu(t)-\log(S_t)] dt + \sigma dW_t,
\label{dejongDyn}
\end{eqnarray}
where $W$ is a standard Brownian motion, $\mu$ is a time-dependent
 parameter, calibrated to the initial futures curve $(F(0,T))_{T\geq 0}$, 
provided by the market;
the mean  reversion parameter $\kappa$ and the volatility $\sigma$ are 
two positive constants. 
As pointed out by \cite{bjerksund2011}, this framework has several drawbacks with respect to the goal of capturing the value of the gas storage. In fact, the
 calibration of the time-varying function $\mu(t)$ is, as expected, quite unstable and gives unrealistic sensitivity of the spot dynamics, and hence of the gas storage value with respect to the initial futures curve. Also, more importantly, this spot modeling does not take into account the futures market and the possibility of trading strategies on futures contracts, while it is well known that spot price is strongly correlated with the evolution of the futures curve, especially the short-term contracts (prompt contract). The futures curve is only used 
 as an initial input to calibrate the parameter $\mu$, but no dynamics for the futures curve is assumed.
Indeed,  modeling  the futures curve is important in order 
to formulate hedging strategies based on futures contracts. \\

To take into account the correlation between futures curve and spot, 
it seems reasonable to introduce 
 models that combine the spot and prompt price in the same dynamics. 
Finally, \ref{dejongDyn} is too poor to describe
stylized facts about  the spot price, as seasonality and the presence of spikes, which are the main sources of the gas storage value.
 An enhancement of this model is proposed by \cite{parsons2012},
 who considers the following two-factor mean-reverting model:
\begin{eqnarray}
\frac{dS_t}{S_t} &=& a [\mu(t)+\log(L_t)-\log(S_t)] dt + \sigma_{S,t} dW_t,\\
\frac{dL_t}{L_t} &=& b [\log(\mathcal{L})-\log(L_t)] dt + \sigma_{L,t} dZ_t,
\label{parsons}
\end{eqnarray}
where $S$ is the spot price described by mean-reversion dynamics
whose  long-run mean has a  stochastic component $L$ and deterministic
value  $\shl$. \\

While this model is more realistic than the one factor model, it still
 suffers from the instability of the deterministic function $\mu$, and 
still does not include the possibility of spikes in the spot price. 
The author defines the futures contract price as the
 expectation of the spot price at maturity date $T$. We emphasize
 that this definition implies
that natural gas is delivered at the futures expiration $T$; however
 in reality the delivery period is spread over a whole calendar month. \\ 

A second way to model the spot process is to consider it as the limit of  the futures
 contract price as time to maturity goes to zero; in particular we have
 $ S_t = \lim\limits_{T\downarrow t} F(t,T)$. This approach was adopted by \cite{warin2010gas};
 the author considers a $n$-factor log-normal dynamics for the futures curve:
\begin{eqnarray}
\frac{dF(t,T)}{F(t,T)} = \Sum_{i=1}^n \sigma_i(t) e^{-a_i(T-t)} dZ^i_t,
\label{warinDyn}
\end{eqnarray}
and by continuity  the spot  is given by $S_t=F(t,t)$. 
In this framework, the author presented a similar algorithm to 
\cite{boogert2008gas}
to estimate  the optimal strategy for the spot; moreover he gives formulae for
 the sensitivities of storage value with respect to futures contracts,
 which enable a hedging strategy to be set up, based on futures, in parallel to 
  the spot optimal trading strategy, which reduces 
 the uncertainty of the realized cash flows.
 This futures-based hedging strategy presents a big advantage, 
compared to the first approach, 
since it increases the manager's chances
 of recovering the storage value
and consequently
the price paid to rent the storage unit.


In conclusion, we have decided to jointly model the futures curve
and the spot prices.
 Indeed, we will formulate a  multi-factor model for the futures curve,
 that includes seasonality of natural gas futures prices, and a dynamics of
 spot price that is correlated to futures curve, more precisely to its short end. We will also incorporate the presence of spikes in the spot prices.

\section{Our modeling framework}

\label{S5}

In Section \ref{stylFact} we discussed the main stylized facts of natural gas prices, which are seasonality and spikes. We believe that the incorporation of these two features is essential in order to monetize these two sources of value. Also, we emphasize that it is crucial to use a modeling framework
 that combines spot and futures curve dynamics, and takes into account
 the existence of a basis between spot and prompt prices. \\

In Section \ref{MFC}, we  will introduce  a two-factors model for the 
futures curve, with a seasonal component for instantaneous volatility.
This parsimonious model has easy-to-interpret parameters and an efficient 
calibration procedure using futures curve historical data.

In Section \ref{spotModel}, we discuss spot price modeling:  we consider two models,
 with a clear relation to the prompt contract. We also include spikes by means of a fast-reverting jump process, similar to a model by \cite{kluge2009}, which was applied to the electricity market.

\subsection{Modeling the futures curve}
\label{MFC}

The first models for  energy futures curves $F(t,T)$ were obtained through conditional expectations of $S_T$ with
respect to the current information at time $t$, where $S$ is the spot price process,
which is indeed modeled, also taking into account  possibly stochastic quantities such as
the convenience yield and the interest rate.
This approach has several drawbacks such as the difficulty of observing or estimating those  quantities
and the problem of fitting the initial curve $F(0,T)$.
\\

 Hence a second stream of models was proposed to directly describe the futures curve, using multi-factor log-normal dynamics.
 For instance, \cite{clewlowOneFactor1999} proposes a one-factor model for the futures curve; this was then extended by 
\cite{clewlowMultiFactor1999} 
to a multi-factor setting. A two-factor version of this model can be expressed  as 
\begin{eqnarray*}
\dfrac{d F(t, T)}{F(t, T)} = e^{-\lambda (T-t)} \sigma_{ST} d W^S_t + \sigma_{LT} d W^L_t ,
\end{eqnarray*} where $\lambda$, $\sigma_{ST}$ and $\sigma_{LT}$ are positive constants, and $W^S$ and $W^L$ are two correlated Brownian motions.
This model has the advantage of exactly fitting the initial futures curve, and the dependence of the volatility on the maturity parameter, i.e. it is of term-structure type; however it does not take
into account the essential seasonality feature. Note that this model is an adaptation of the well-known \cite{gabillon1991} model, 
originally proposed for spot prices. Our framework slightly modifies previous models,  adding a seasonality component and
introducing  parameters that  have an economical significance. \\

We will call it Seasonal Gabillon two-factor model. It is formulated as
\begin{eqnarray}
\dfrac{d F(t, T)}{F(t, T)} = e^{-\lambda (T-t)} \phi(t) \sigma_S d W^S_t + 
(1-e^{-\lambda (T-t)}) \sigma_L d W^L_t ,
\label{gabillon}
\end{eqnarray}
where $W^S$ and $W^L$ are two correlated Brownian motions, with $d\langle W^L,W^S \rangle_t=\rho dt$. The letters $L$ and $S$ 
stand respectively for Long term and  Short term; $\lambda$, $\sigma_{S}$ and $\sigma_{L}$ are positive constants.
The function $ \phi(t) = 1 + \mu_1 cos(2\pi(t-t_1)) + \mu_2 cos(4\pi(t-t_2))$ weights instantaneous volatility with a periodic behaviour. It takes into account the winter seasonal peaks (resp. the secondary summer peak)
 by taking for example $t_1$ equal to January (resp.  with $t_2$ equal to August).
 The coefficients $\mu_1$ and $\mu_2$  quantify  the winter and summer seasonality contribution in the volatility:
 we expect the winter parameter $\mu_1$ to be often larger, in absolute value, than the summer parameter $\mu_2$. \\

Our model constitutes an efficient  framework,  whose parameters  
are economically meaningful. Indeed,
the parameters $\sigma_L$  and   $\sigma_S$
play the role of a 'long-term', respectively
 'short-term' volatility.
 Note that even if the model is expressed with a continuous set of maturities, in the real world we only have access to a finite number
 of maturities, for example, monthly spaced futures contracts.

In the next section
we give more details about the meaning of each parameter and  their estimation, using historical data of futures prices.

\subsubsection{Model estimation}

Many of the model parameters are almost observable,
if we have sufficient historical data of futures curves at hand. In fact, $\sigma_S$ and $\sigma_L$ could be approximated by the volatility of short and long-dated continuous futures contracts, and $\rho$ by their empirical correlation.\\

In fact, for $T \to \infty$, we can formally write $ \dfrac{d F(t, T)}{F(t, T)} \simeq \sigma_L d W^L_t$, so a good approximation for the long-term volatility is $$ \sigma_L^2 \simeq \dfrac{1}{m-1} \Sum_{i=1}^{m} (\dfrac{z_{t_i}^L}{\sqrt{\Delta t_i}} - \bar{\mu}^L)^2 ,$$ where $z_t^L$ is the log-return of a constant maturity long-dated contract, four years for example, and $\bar{\mu}^L = \dfrac{1}{m} \Sum_{i=1}^{m} \dfrac{z_{t_i}^L}{\sqrt{\Delta t_i}}$.\\

On the other hand, for small times to maturity, i.e.
 $T - t \to 0$, we can ignore the long-term noise effect, and write $ \dfrac{d F(t, T)}{F(t, T)} \simeq \sigma_S d W^S_t$, 
so that a good proxy for the spot volatility is the volatility of the rolling prompt contract, i.e. the contract with the nearest maturity $$ \sigma_S^2 \simeq \dfrac{1}{m-1} \Sum_{i=1}^{m} (\dfrac{z_{t_i}^P}{\sqrt{\Delta t_i}} - \bar{\mu}^P)^2 ,$$ where $z_t^P$ is the log-return of a prompt futures contracts and $\bar{\mu}^P = \dfrac{1}{m} \Sum_{i=1}^{m} \dfrac{z_{t_i}^P}{\sqrt{\Delta t_i}}$.\\

We can also give an initial estimate for the correlation parameter $\rho$ as $$ \rho \simeq \dfrac{1}{m-1}  \frac{\Sum_{i=1}^{m} (\dfrac{z_{t_i}^P}{\sqrt{\Delta t_i}} - \bar{\mu}^P) (\dfrac{z_{t_i}^L}{\sqrt{\Delta t_i}} - \bar{\mu}^L)}{\sigma_S \sigma_L}.$$

These rough estimates could be used directly, or as input parameters for a more rigorous statistical estimation procedure. For example, we can use the maximum 
likelihood method. For that, suppose we have a time series over dates $t_1, \ldots, t_m$ of futures prices maturing at $T_1, ..., T_n$. We  
denote $z_t, t  = t_i, i \in \{0,\ldots,t_{m-1}\}$ 
 the vector of price returns,  $\Delta t$ being the corresponding step $t_{i+1} - t_i$ and $\theta$ is the model parameters 
vector $\theta = (\lambda, \mu_1, \mu_2, \sigma_S, \sigma_L, \rho)$,
we have
\[ z_t =   \left( \begin{array}{c}
\dfrac{\Delta F(t,T_1)}{F(t,T_1)}\\
. \\
. \\
. \\
\dfrac{\Delta F(t,T_n)}{F(t,T_n)} \\
\end{array} \right) ,\quad
H_t = \sqrt{\Delta t} \left( \begin{array}{cc}
e^{-\lambda(T_1-t)}\phi(t)\sigma_S, & (1-e^{-\lambda(T_1-t)}) \sigma_L \\
. & . \\
. & . \\
. & . \\
e^{-\lambda(T_n-t)}\phi(t)\sigma_S, & (1-e^{-\lambda(T_n-t)}) \sigma_L
\end{array} \right),
\]
where $ \Delta F(t,T_1) =  F(t+ \Delta t,T_1)-F(t,T_1)$. Then an Euler discretization of the SDE \eqref{gabillon} gives the equation
$$
z_t = H_t x_t, \ t \in \{t_1, \ldots, t_m \},
$$
where $(x_{t_i})$ are independents Gaussian 2-d vectors such that
$$x_{t_i} \ \mathbb{s} \ \mathcal{N}(0, \Sigma), \ 1 \le i \le m,$$
where \[ \Sigma = \left( \begin{array}{cc}
1 & \rho  \\
\rho  & 1
\end{array} \right).\]
The likelihood maximization could then be written as the minimization of the function
$$
L(x_{t_1}, x_{t_2}, ..., x_{t_m} | \theta) =  \dfrac{1}{m} \Sum_{i=1}^{m} log(det(\Sigma)) + x_{t_i}^T \Sigma^{-1} x_{t_i},
$$
and the $x_t, t \in \{t_1, \ldots, t_m\} $ are given by $ z_t = H_t x_t$, i.e. $$x_t = (H_t^T H_t)^{-1} H_t^T z_t.$$
So, the model estimation procedure is equivalent to the following minimization problem
\begin{equation}
\left\{
\begin{tabular}{l}
    $\displaystyle   \min \; L(x_{t_1}, x_{t_2}, ..., x_{t_m} | \theta) = log(det(\Sigma)) + \dfrac{1}{m} \Sum_{i=1}^{m} x_{t_i}^T \Sigma^{-1} x_{t_i}$ \\
    $\theta = (\lambda, \mu_1, \mu_2, \sigma_S, \sigma_L, \rho).$
\end{tabular}
\right.
\label{futEstimation}
\end{equation}

To illustrate, we apply this   estimation procedure, to historical data based on a 1997-2007 futures curves history. As mentioned,
 the estimation problem \eqref{futEstimation} is solved using an optimization algorithm, with the rough estimates of $\sigma_S$, $\sigma_L$ 
and $\rho$ as initial point for the algorithm. We report in Table \ref{futureModelParams} the estimated parameters of the futures curve model.\\
 
\begin{table}[h]
   \centering
	\begin{tabular}{|c|c|c|}
\hline
Parameter & Value & Confidence interval\\
\hline
$\sigma_S$ &0.4580& [0.4462,0.4698] \\
$\sigma_L$ &0.1655& [0.1617,0.1694] \\
$\lambda$ &0.7896& [0.7518,0.8274] \\
$\mu_1$ &0.0246& [-0.0015,0.0507] \\
$\mu_2$ &0.0038& [-0.0218,0.0294] \\
$\rho$ &0.4113& [0.3737,0.4488] \\
\hline
	\end{tabular}
	\caption {Estimated parameters using 1997-2007 futures curves history.}
	\label{futureModelParams}
\end{table}

As expected, the short-term volatility is larger than the long-term volatility, which is a common feature in energy futures curve dynamics, and the winter contribution $\mu_1$ in the seasonality component is larger than summer contribution $\mu_2$.

\subsection{Modeling spot price}
\label{spotModel}

The main characteristic of the modeling framework introduced by \cite{warin2010gas} is the assumption that the futures price maturing at $T$, $F(t,T)$,
 converges to the spot price $S_t$ when the time to maturity is close 
to zero; as we mentioned, this hypothesis  is not realistic, for the simple reason that
a futures contract $F(\cdot, T)$  delivery 
does not take place at the fixed expiration day  $T$ 
 but spreads out over a period of one month. Besides, the delivery point of futures and spot might be different.
 We also note  that the spot is subject to additional noise compared to futures contracts, as unpredicted weather changes or
 technical incidents have a larger impact on the spot than on the futures itself. \\

All these considerations suggest the spot should be considered as a separate, but correlated, stochastic process when in general
  $$ S_t \neq \displaystyle \lim_{T \to t} F(t,T). $$

A model in that sense was proposed by \cite{gray2010}, where  the logarithm of the spot is a mean reverting process, whose  mean-reversion
 level is a stochastic process equal to the prompt price. For a family of maturities $(T_i)_i$,   the futures contract $ F(t,T_i)$
is a log-normal process fulfilling  
$$
 \dfrac{d F(t,T_i)}{F(t,T_i)} = \sigma(t,T_i) d W_t$$ and the spot price $S_t$ evolves along
\begin{equation} \label{EGray}
d \log(S_t) = (\theta_t + a \log(P_t) - a \log(S_t))dt + \sigma^S_t d B_t,
\end{equation}
where $B$ and $W$ are two correlated Brownian motions, and for the current date $t$, $P_t$ denotes the prompt price, i.e. $$P_t=F(t,T_i) \quad \text{for} \quad T_{i-1} \leq t < T_i.$$

In our opinion it is crucial to incorporate futures curve dynamics into the modeling of the spot prices,
for instance a dynamics relating the spot and prompt futures price.
Indeed, as shown by the historical paths of spot 
and prompt prices in  Figure \ref{fig:SpotPromptHistorical}, the two processes are closely related. In fact they seem to move very often in the same direction, 
with some occasional dislocations of spot and prompt prices. \\

In what fallows we will study two spot models, in relation to our futures curve model. These will be stated in discrete time.


\subsubsection{Spot model 1}

Our first spot model is similar to \eqref{EGray}, which was introduced by 
\cite{gray2010}. Its dynamics, based on the spot log-return 
$y_t= \log(S_t/S_{t-1})$,  
is given by
\begin{equation}\label{spotModel1}
\log(S_t/S_{t-1}) = a_1 + a_2 \log(P_{t-1}/S_{t-1}) + a_3 \log(P_{t}/P_{t-1}) + \epsilon_t
\end{equation}
where $(\epsilon_t)$ is a Garch$(p, q)$ process
and  again $P$ is the prompt price. \\

Recall that a Garch(p, q) process $\epsilon$ verifies an autoregressive
moving-average equation for the error variance
\begin{eqnarray}
\label{garch}
\epsilon_t &=& \sigma_t z_t \text{ , where} \nonumber \\ 
\sigma_t^2 &=& \kappa + \Sum_{i=1}^{p} \gamma_i \sigma_{t-i}^2 + \Sum_{i=1}^{q} \alpha_i \epsilon_{t-i}^2
\end{eqnarray}
and $z$ a white noise.\\

This model intends to capture both the heteroscedasticity of the natural gas spot price and the correlation between spot price and prompt futures price.
In fact, similarly to \eqref{EGray},  we remark here that in our
 dynamics \eqref{spotModel1},  the spot price is mean reverting
around a stochastic level equal to prompt price. 
On the other hand, the prompt log return is a supplementary 
 explanatory variable of the spot log return. 
We recall that our futures model \eqref{gabillon} incorporates seasonality in the futures curve dynamics; this implies that the spot dynamics itself, 
by means of the prompt price, has an implied seasonality component. This
allows us to avoid the addition of some seasonal function into
 the spot dynamics \eqref{spotModel1}.

\subsubsection{Spot model 2}

The second spot model we propose is based on the series of spot-prompt spread  $y_t= \frac{S_t - P_t}{P_t}$. In fact, we model the spot-prompt spread, through
the so-called  front-back spread as  regression variable.
In other words, we write 
\begin{eqnarray}
\frac{S_t - P_t}{P_t} = a_1 + a_2 \frac{S_{t-1} - P_{t-1}}{P_{t-1}} + a_3 \frac{P_{t-1} - B_{t-1}}{B_{t-1}} + \epsilon_t,
\label{spotModel2}
\end{eqnarray}
where $B_t$ is the price of the second nearby futures (also known as the back contract) and $\epsilon$
 is a Garch(p, q) process. \\

\eqref{spotModel2} has the advantage of directly handling the spread between spot and prompt prices, which is probably a good indicator of the decisions
  to be made in gas storage management. In fact, a large positive spread value would possibly induce the decision to withdraw gas, while the inverse would motivate a gas injection. Also, as we pointed out in the introduction, the narrowing of the seasonal spread in the futures curve during last years has diminished the intrinsic value of gas storage units. Consequently, almost all the storage value is now concentrated in the extrinsic value, which is
 heavily dependent on the spot-prompt spread.\\


\subsubsection{Spikes modeling}

In Section \ref{stylFact}, we showed that natural gas prices have two special characteristics: seasonality and  presence of spikes. The first feature
 (seasonality), is included in the spot dynamics through the prompt (and the
 second nearby) futures contract. In fact, the futures curve dynamics \eqref{gabillon} already has a seasonal component, so we have
 chosen not to add a supplementary seasonal part in the spot dynamics.
 On the other hand, spikes are included in the spot model
 via a 
 jump process.
 These large and rapidly absorbed jumps are an essential feature of the spot, since they can be source of value for gas storage and
 they can be monetized if injection/withdrawal rates are high enough.
\\
 
Indeed, we describe the spikes as a fast mean-reverting (jump) process,
already introduced
 by \cite{kluge2009}, in the framework of  the electricity market. 
These authors proposed a spot model for the power price that incorporates the presence of spikes via 
a process $Y$, being the solution of the equation
\begin{eqnarray}
d Y_t &=& -\beta Y_{t-} dt + d Z_t, \quad   Y_0 = 0,
\label{spike}
\end{eqnarray}
where $Z$ is a compound Poisson process of the type $Z_t = \Sum_{i=1}^{N_t}  J_i$, $(N_t)$ is a Poisson process with intensity $\lambda$ 
and $(J_i)_{i\in \N}$ is a family of independent identically distributed (iid) variables representing the jump size. Furthermore ($N_t$) and 
($J_i$) are supposed to be mutually independent. The process $Y$ can be written explicitly as
\begin{equation} \label{SolSpikes}
  Y_t = Y_0 e^{-\beta t} + \Sum_{i=1}^{N_t} e^{-\beta (t - \tau_i)} J_i.
\end{equation}
We recall that the spot model is directly expressed as a discrete time process, indexed on the grid $(t_i)$ introduced in Section \ref{S3}. For that reason $Y$ will be restricted to the same time  grid. \\

 Choosing a high value for the mean-reversion parameter $\beta$ forces the jump process $Y$ to revert very quickly to zero
 after the jump times $\tau_i$, which constitutes a desired feature for natural gas spikes. In fact, the jumps in natural gas spot prices are
 rapidly absorbed, thanks to the storage capacities available in the market.\\

We emphasize that the two models \eqref{spotModel1} and \eqref{spotModel2} alone do not take into account the possibility of sudden spikes in the spot price.
The process $Y$ will be indeed incorporated 
 into the dynamics in \eqref{spotModel1} and \eqref{spotModel2},
 by multiplying the spot process by the process $\exp(Y_t)$, i.e.
$$ \tilde{S}_t =  \exp(Y_t) S_t.$$

As we noted in Section \ref{stylFact}, the natural gas spikes are clearly distinguished by their signs.
 In fact, positive spikes, due to unpredicted weather changes, occur exclusively during the winter and summer months. On the other hand, 
negative spikes, generally caused by poor anticipation of the storage capacities of the market, happen mostly during shoulder 
months like October and November. 
This motivates a separate modeling for these two categories of spikes.
 We will consider two processes $Y^+$ and $Y^-$ for positive and negative spikes, each one verifying a slightly modified version of the equation \eqref{SolSpikes}, as follows:
\begin{equation}
\label{SolSpikesSeas}
  Y^+_t = \Sum_{i=1}^{N_t} e^{-\beta (t - \tau_i)} J_i 
\mathbbm{1}_{\tau_i \in I^+},
\end{equation}
where $I^+$ (resp. $I^-$)
represents the positive (resp. negative) spikes occurring period, i.e. winter and summer
(resp. shoulder months),  as we observed in Section \ref{stylFact}.
\\

Consequently, the spot process that we consider for our gas storage valuation is
\begin{equation}
\tilde{S}_t =  \exp(Y^+_t + Y^-_t) S_t.
\label{spotWithSpikes}
\end{equation}
Finally \eqref{spotWithSpikes} will have all the desired properties: 
it includes seasonality by  relating the futures curve to the spot dynamics 
 and it allows the presence of positive and negative spikes, each one generated by a separate jump process $Y^+$ and $Y^-$.\\

\subsubsection{Model estimation}

Similarly to the model for futures, we use historical data for spot and 
futures prices. The parameters estimation for the two spot dynamics 
\eqref{spotModel1}, \eqref{spotModel2}
proposed above is based on regression techniques and the classic estimation 
procedure for Garch processes.
Similarly to  \cite{kluge2009}, 
 we also use the likelihood method to estimate spike process parameters,
 after filtering the underlying time series to extract the jumps.
Note that the coefficient $\beta$ is heuristically fixed. \\
\begin{table}[ht]
   \centering
	\begin{tabular}{|c|c|}
		\hline
		Regression parameters & Value\\
		\hline
		$a_1$ &  -0.0044\\
		$a_2$ &  0.2622 \\
		$a_3$ &  0.4467\\
		\hline
		Garch(1,1) parameters & Value\\
		\hline
		$\kappa$ & 1.6928e-005\\
		$\gamma_1$ & 0.8764 \\
		$\alpha_1$ & 0.1138 \\
		\hline
		Spike process $Y^+$ & \\
		\hline
		$\beta$ & 300 \\
		$\lambda$ & 0.8331\\
		Jump Law & $\mathcal{N}(0.2579,0.3910)$\\
		\hline
		Spike process $Y^-$ & \\
		\hline
		$\beta$ & 300 \\
		$\lambda$ & 2.9488\\
		Jump Law & $\mathcal{N}(-0.7624,0.6402)$\\
		\hline
	\end{tabular}
	\caption {\label{spotParams} Spot model 1 parameters using 1997-2007 data}
\end{table}

An analysis of the spot and futures historical data shows that a Garch(p, q) process of order $p=1$ and $q=1$ is sufficient, using higher order being of insignificant impact. As mentioned before, we use a large value for the spike reversion parameter $\beta$. \\

To illustrate, the estimation procedure of the parameters of the spot model 1, using a Garch(1,1) process and a 1997-2007 history of spot and futures prices, yields the parameters in Table \ref{spotParams}.

\section{Numerical results}
\label{SNumRes}

In this section we use our futures-spot modeling to value a storage contract, and compare it with the intrinsic valuation method. For this we will concentrate on  fast and slow storages, which constitute two realistic cases. A fast gas storage has high injection/withdrawal rates, so that it  can be filled in general within a month, but it has low gas capacity: salt caverns are a common example of high deliverability storage units. Slow gas storages are in general large depleted oil/gas fields, or aquifers, so they have very large gas capacities, but they suffer from low injection/withdrawal rates (see Appendix \ref{appA}
 for more details).
\\

We will consider a fast and a slow storage unit with the  characteristics described  in Table \ref{storageUnit}, where for simplicity, all the quantities are expressed in $10^6$ MMBtu\footnote{This energy unit can be naturally converted into 
a volume, under standard conditions for temperature and pressure.}, 
while the storage values are expressed in $\$ $ million. This means that fast storage takes 25 days to fill, and almost 17 days to empty, while slow storage needs 125 days to be completely filled and 83 days to be completely  emptied. We also consider null injection/withdrawal costs.\\

\begin{table}[ht] 
   \centering
	\begin{tabular}{|c|c|c|}
		\hline
		& Fast storage & Slow storage \\
		\hline
		Total capacity & $100$ & $100$  \\
		\hline
		Injection rate & $4$ per day & $0.8$ per day \\
		\hline
		Withdrawal rate & $6$ per day & $1.2$ per day \\
		\hline
		Initial gas volume & 0 & 0 \\
		\hline
		Final gas volume & 0 & 0 \\
		\hline
		Lease duration & 1 year & 1 year \\
		\hline
	\end{tabular}
	\caption {\label{storageUnit} Gas storage characteristics (fast and slow units)}
\end{table}

\begin{figure} 
\setlength\figureheight{3.5cm} 
\setlength\figurewidth{15cm}
\input{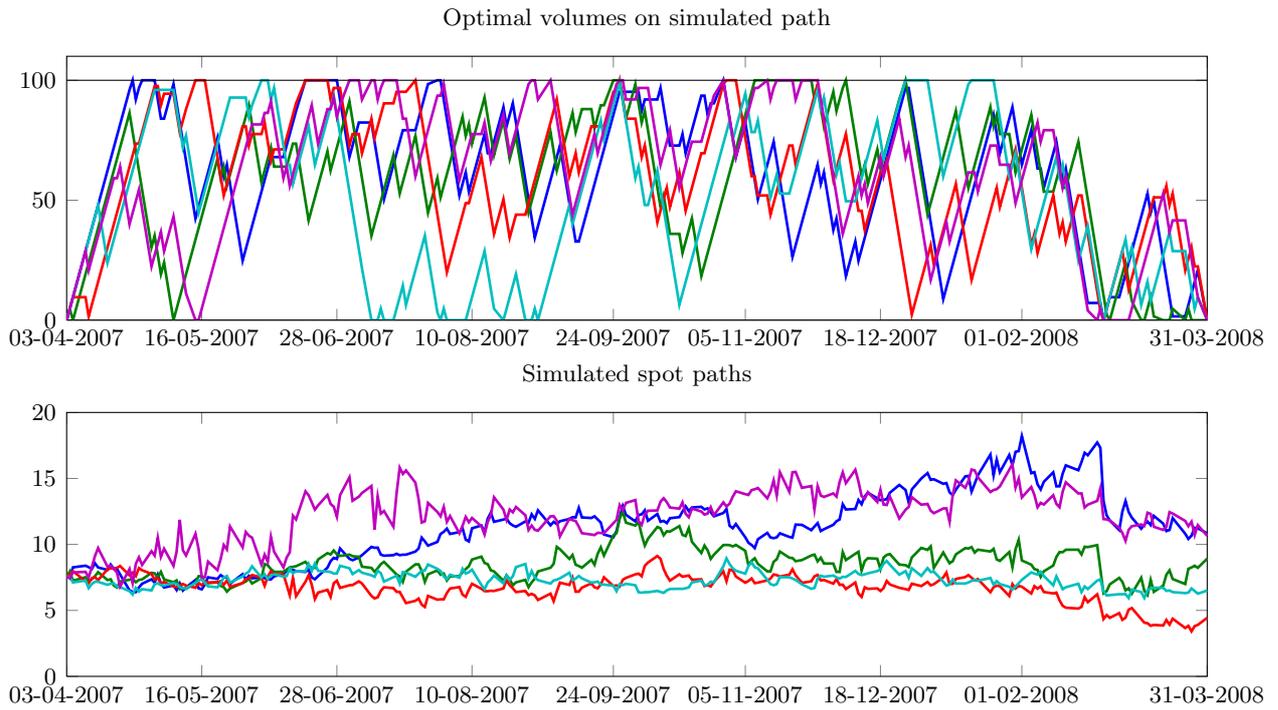} 
\caption{Simulated spot paths and optimal volumes (fast storage)} 
\label{fig:OptimalVolumesFastStorageSim} 
\end{figure}

\begin{figure} 
\setlength\figureheight{3.5cm} 
\setlength\figurewidth{15cm}
\input{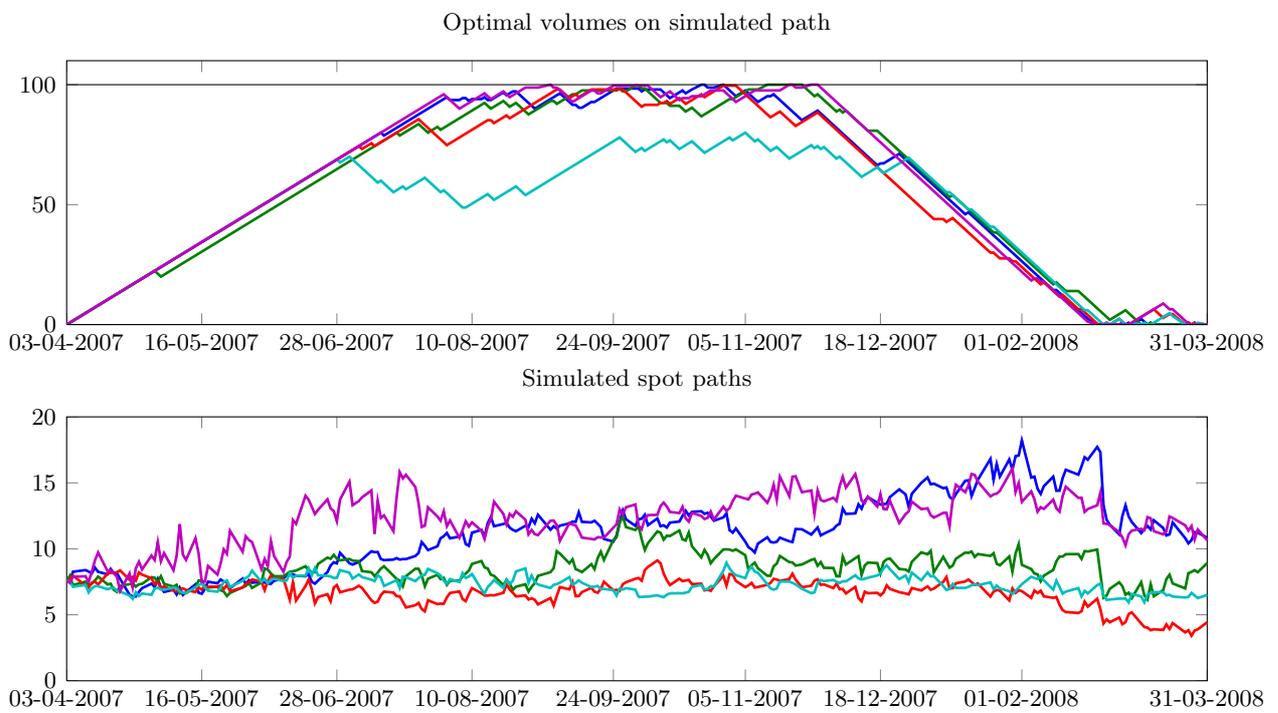} 
\caption{Simulated spot paths and optimal volumes (slow storage)} 
\label{fig:OptimalVolumesSlowStorageSim} 
\end{figure}

The experiments were run using the Matlab software, with 5000 simulations for the Monte Carlo method, and we used independent paths for the  backward and forward phases, in the Longstaff$\And$Schwartz algorithm (see Section \ref{SDPE}). First, we simulate a set of spot and futures paths, then we apply the dynamic programming algorithm \eqref{dyn_vol} to estimate the optimal spot strategy; in parallel we evaluate the hedging strategy, based on futures contracts, according either to \eqref{EDelta1} or \eqref{EDelta2}. We then re-simulate a new set of spot and futures paths, independent from the paths used in the preceding backward phase, and we apply the estimated optimal spot strategy, combined with the futures hedging strategy, to the new trajectories.
 We store the cumulative cash flows $\text{Wealth}_{\text{spot+futures}}(u^{\star})$ resulting from these physical and financial operations for each sample path, and we compute the empirical mean and standard deviations of those cash flows. The mean of the 
cumulative wealth gives an estimate of the extrinsic value $J^{\star}$ of the gas storage, given in \eqref{maxPb},  while the standard deviation is an indicator of the dispersion of the realized cash flows around the extrinsic value. 
We emphasize that the empirical mean estimates the cash flow generated by the optimal strategy, while the empirical standard deviation gives an indicator of the variance reduction obtained through the financial hedging strategy. A lower standard deviation means that the manager will face less uncertainty on a single realization of spot and futures prices. Numerical results will confirm that the hedging strategy indeed allows for a significant variance reduction of the cumulative cash flows. An example of the outputs of the above valuation procedure is represented in
 Figure \ref{fig:OptimalVolumesFastStorageSim} (fast storage) 
and \ref{fig:OptimalVolumesSlowStorageSim} (slow storage) 
 by samples of simulated spot trajectories, with the corresponding optimal gas volumes in the unit, for a contract starting in April $2007$. Note that different colors correspond to different
simulated spot trajectories.
\\

\begin{figure} 
\setlength\figureheight{3.5cm} 
\setlength\figurewidth{15cm}
\input{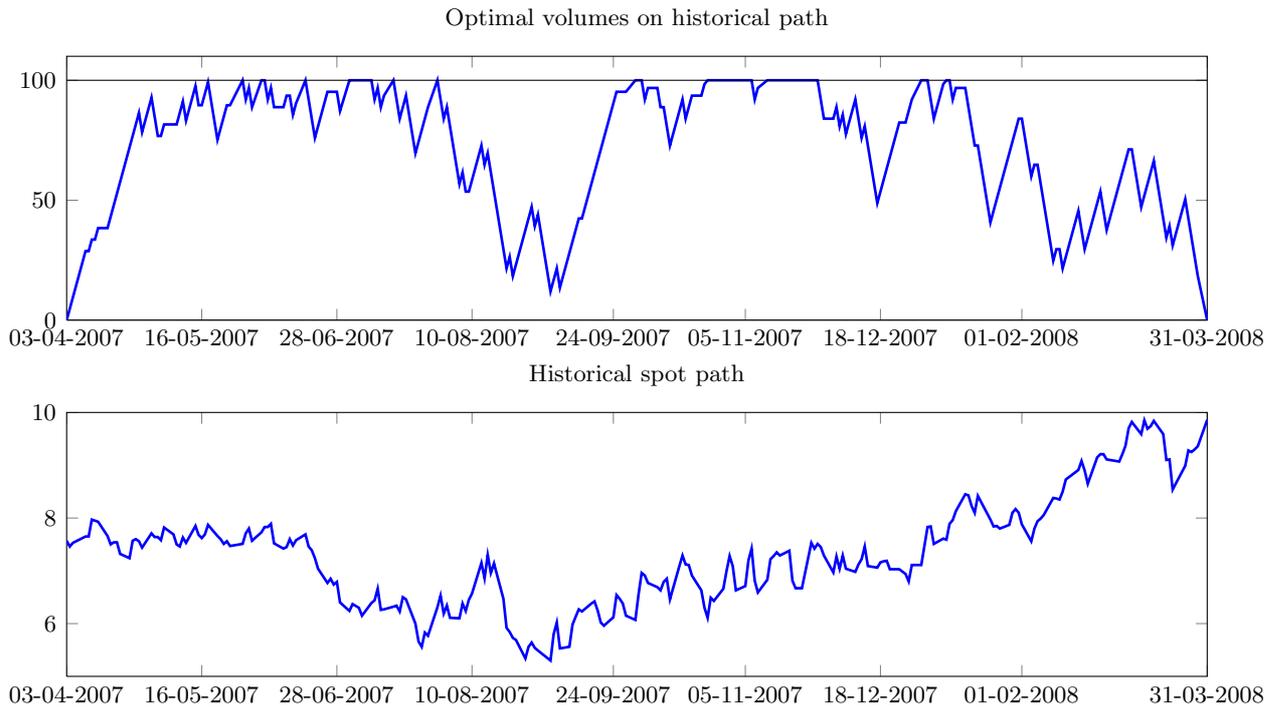} 
\caption{Historical spot path and optimal volumes (fast storage)} 
\label{fig:OptimalVolumesFastStorageHist} 
\end{figure}

\begin{figure} 
\setlength\figureheight{3.5cm} 
\setlength\figurewidth{15cm}
\input{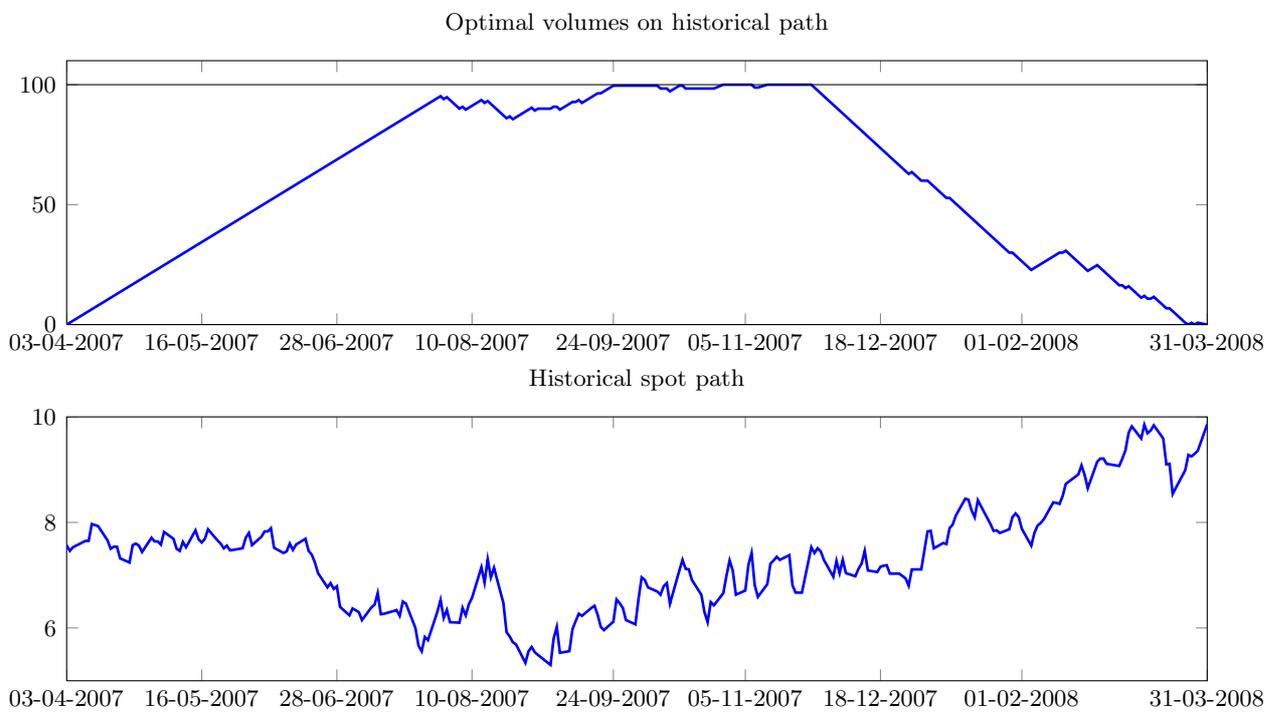} 
\caption{Historical spot path and optimal volumes (slow storage)} 
\label{fig:OptimalVolumesSlowStorageHist} 
\end{figure}

Note also that the analysis described above depends on the choice of the model,
 because the backward and the forward phases are executed on the sample 
paths generated by the model itself. In order to make the comparison less model-dependent, we calculate  the cumulative cash flows  of the estimated optimal
 strategy, based on  spot and futures {\it historical paths}.
 For this reason, we will consider a series of spot and futures curve data from 2003 to 2012, and split it into periods of one year: the storage lease contracts  specified in Table \ref{storageUnit} start in April each year, for a one-year period. We run the optimal strategy  
obtained in the backward phase (for the corresponding storage duration) on the spot and futures  historical paths for the related period. This constitutes a real case test for the optimal strategy and corroborates the relevance of the spot modeling, since it provides the profit that would have been accumulated by the storage manager in a realized path.
 Figures \ref{fig:OptimalVolumesFastStorageHist} and \ref{fig:OptimalVolumesSlowStorageHist} represent the historical spot path realized during the contract period (for both slow and fast units) from April $2007$ to April $2008$, and the natural gas volumes resulting from the optimal strategy computed on simulated paths (see Figures \ref{fig:OptimalVolumesFastStorageSim} and \ref{fig:OptimalVolumesSlowStorageSim} for examples of these simulated paths). \\

We summarize the results of the valuation algorithm for each period in Tables \ref{testFast} and  \ref{testSlow} in fast and slow storage cases, 
when the spot paths are generated according to the spot model 2. proposed in \eqref{spotModel2}. The tables report, for each period, the intrinsic value (IV) (see \eqref{IntrinsicValue} in Appendix \ref{appB}) and the estimate of the extrinsic value (EV) on simulated paths and historical paths realized during the current period. The last two columns show the standard deviation of the simulated cash flows under the optimal strategy.\\

We expect that the extrinsic spot-based strategy will give a larger value than the intrinsic physical futures-based strategy, while our financial hedging strategy is supposed to reduce the uncertainty of gas storage cash flows. For example, the fast storage contract starting in April $2007$ has an intrinsic value of $\$ 222.9689 \ 10^6$ while the spot-based strategy gives an extrinsic value of $\$ 697.0003 \ 10^6$.  
As expected, the extrinsic strategy allows better financial exploitation of the rights (without obligation) of injection/withdrawal natural gas compared to the conservative intrinsic strategy. In other words, the extrinsic strategy allows better extraction of the optionality of storage. We also note that the hedging strategy yields a significant empirical variance reduction of the cumulative cash flows from $\$340.2193 \  10^6$ to $\$190.8546 \ 10^6$. 
On the other hand, the intrinsic value of slow storage is equal 
to $\$195.5517 \ 10^6$, while the spot-based strategy captures a larger optionality value of $\$251.0064 \ 10^6$. Similarly to fast storage, the financial hedging strategy allows an important variance reduction from
 $\$232.7825 \ 10^6$ to $\$28.0414 \ 10^6$. \\

Previous observations about the 2007 contract remain valid for the other test periods; indeed the intrinsic futures strategy is always out-performed by the extrinsic spot-based strategy, in both simulated and historical paths. The historical backtesting over the period 2003-2012 shows that the extrinsic strategy allows for better extraction of storage unit optionality, with a 
ratio of extrinsic value to intrinsic value as high as $500\%$ for a fast storage unit. 
This performance of the extrinsic strategy is less significant in the case of slow
 storage unit, with a ratio up to $100\%$. This is due to limitations
 in the deliverability of slow storage, since the storage manager 
is not able to profit completely from the gas price volatility and 
cannot respond rapidly to favorable price movements.
 On the other hand, hedging with financial instruments provides a significant reduction  of the cumulative cash flows uncertainty. In fact, the last two columns of Tables \ref{testFast} and \ref{testSlow} show a standard deviation reduction factor of up to $10$, with better performance for slow storage. This gives the storage manager more insurance to recover a large percentage of the value of the storage contract.\\

\begin{remark} \label{RM1M2}
\begin{enumerate}
\item In Section \ref{hedging}, we presented two heuristic hedging strategies, \eqref{EDelta1} and \eqref{EDelta2}, based on financial futures contracts. The numerical tests that we have conducted show that the hedging strategy
 defined by \eqref{EDelta2} gives better results in the variance reduction of the simulated cash flows under the optimal strategy; in addition,  in the historical backtesting,
\eqref{EDelta2} renders a better cumulative wealth performance  
than \eqref{EDelta1}. We emphasize that 
we have only reported about the better performing  
hedging strategy \eqref{EDelta2}.
\item
We also note that the historical intrinsic value of the gas storage attains a peak in 2006, and shows a clear decreasing effect afterwards. This can be intuitively explained observing the futures curve samples in Figure \ref{fig:FutureCurveNG}: in 2006,  the seasonal spreads were very pronounced, while they were quite small in 2011. 
\end{enumerate}
\end{remark}

\begin{table}[H]
   \centering
	\begin{tabular}{|c|c|c|c|c|c|c|}
		\hline
		 &\multicolumn{2}{|c|}{Simulated paths test} & \multicolumn{2}{|c|}{Historical path test} & \multicolumn{2}{|c|}{Standard deviation}\\
		\hline
		Starting Date & IV & EV & IV & EV & Without hedge & With hedge\\
		\hline
			2003-Apr &39.9542 &337.7276 &42.6441 &184.1178 &189.5820 &119.3606\\
			2004-Apr &63.0335 &395.6198 &63.6000 &347.2736 &213.1796 &126.3763\\
			2005-Apr &115.2008 &592.0854 &112.0473 &528.6510 &306.3232 &179.4792\\
			2006-Apr &371.1724 &860.9714 &416.3992 &616.2357 &390.0864 &194.9693\\
			2007-Apr &222.9689 &697.0003 &241.8000 &399.7347 &340.2193 &190.8546\\
			2008-Apr &119.5200 &674.6745 &129.6000 &427.9652 &359.7650 &210.6817\\
			2009-Apr &204.6539 &459.5531 &205.9000 &302.6958 &203.2847 &100.9753\\
			2010-Apr &144.1958 &420.2250 &153.7000 &259.1776 &202.1802 &112.6989\\
			2011-Apr &86.5488 &352.7785 &92.9000 &134.7794 &190.0749 &102.9312\\
			2012-Apr &125.8968 &272.5376 &130.2000 &215.9591 &118.0606 &55.4645\\
		\hline
	\end{tabular}
	\caption {\label{testFast} Fast gas storage valuation (under spot model 2 \eqref{spotModel2})}
\end{table}

\begin{table}[H]
   \centering
	\begin{tabular}{|c|c|c|c|c|c|c|}
		\hline
		 &\multicolumn{2}{|c|}{Simulated paths test} & \multicolumn{2}{|c|}{Historical paths test} & \multicolumn{2}{|c|}{Standard deviation}\\
		\hline
		Starting Date & IV & EV & IV & EV & Without hedge & With hedge\\
		\hline
			2003-Apr &24.6556 &67.5795 &26.0563 &16.7890 &83.9382 &18.5218\\
			2004-Apr &45.2183 &91.1053 &44.6833 &53.8389 &119.8064 &20.1098\\
			2005-Apr &93.6136 &157.9486 &92.2304 &146.3097 &189.6219 &28.0376\\
			2006-Apr &333.1988 &386.4656 &333.1972 &356.0564 &282.2993 &29.6749\\
			2007-Apr &195.5517 &251.0064 &195.3024 &221.4466 &232.7825 &28.0414\\
			2008-Apr &96.8824 &169.6740 &98.5936 &141.5038 &216.1477 &32.6633\\
			2009-Apr &180.5010 &206.6439 &180.4980 &210.8117 &148.3145 &14.6222\\
			2010-Apr &122.4013 &152.9924 &122.3784 &128.1330 &140.5389 &16.5936\\
			2011-Apr &68.5264 &104.2509 &68.1356 &72.3083 &118.2672 &18.8294\\
			2012-Apr &107.4493 &122.5703 &107.3928 &110.0214 &86.2897 &8.4167\\
		\hline
	\end{tabular}
	\caption {\label{testSlow} Slow gas storage valuation (under spot model 2 \eqref{spotModel2})}
\end{table}

We conclude from the numerical results presented above that the joint modeling of the natural gas spot price and futures curve is a pertinent framework for the gas storage valuation and hedging problem. It allows the unit manager to better exploit storage optionality by monetizing the spot price volatility and seasonality. Indeed, the historical backtesting shows that the extrinsic value under this modeling always outperforms the classical intrinsic value, even in the case of slow storage. A joint model for the futures curve with its own risk factors is a more realistic framework for spot and futures markets, since it takes into account the seasonality of the futures curve and the non-convergence of the futures price to the spot price, an unrealistic hypothesis that is often made in the literature. This also allows for a more relevant hedging strategy based on futures contracts, and better tracking of the extrinsic value of gas storage in real market conditions.

\section{Model risk}

\label{S7}

As we showed in the introduction, seasonal spreads have become narrower these last years, which leads to a concentration of almost all the value of gas storage in the extrinsic part, based on spot trading. Hence it is very important to look closely into the spot modeling and its effect on storage valuation and hedging. We believe that the uncertainty of storage value is due more to the uncertainty of the spot modeling than the futures modeling, since only spot evolution determines the optimal strategy even though the futures contract prices intervene in spot modeling, see \eqref{spotModel1} and \eqref{spotModel2}. Indeed their main contribution is devoted to variance reduction. In Section \ref{S71} we study the effects of the modeling hypotheses, and sensitivity 
with respect to the model parameters; in Section \ref{S72}, we define a model  risk measure to quantify this dependence, as proposed by \cite{cont2006model}. Before this, in Section \ref{S71},
 we compare the performance of the two spot models proposed in Section \ref{spotModel}, using  historical data.

\subsection{Spot modeling}
\label{S71}

In Section \ref{spotModel}, we proposed two discrete models for the spot price dynamics. The first model, defined in \eqref{spotModel1}, is a discrete version of a mean-reverting model, with a stochastic mean-reversion level equal to the prompt price. The second model, in \eqref{spotModel2}, handles directly the spread between spot and prompt prices, which  could be  a decisive quantity in the optimal strategy. In fact, the unit manager will probably tend to buy and store gas if the spot-prompt spread is negative and withdraw  and sell gas
in the opposite case. Since the seasonality of gas prices has been getting weaker in recent years, the principal source of value for the storage unit is the spot-prompt spread rather than the winter-summer spreads, so the second model \eqref{spotModel2} is expected to give good 
results for storage valuation and hedging. \\

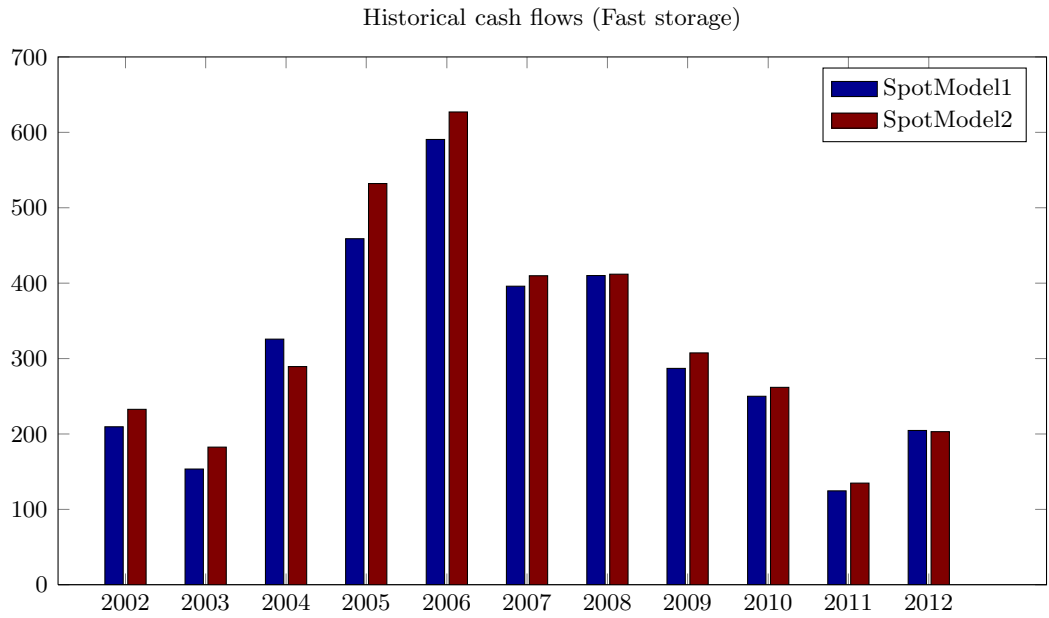
\begin{figure} 
\setlength\figureheight{7cm} 
\setlength\figurewidth{13cm}
%
%
%
%

\definecolor{mycolor1}{rgb}{0,0,0.5625}%

\begin{tikzpicture}

\begin{axis}[%
width=\figurewidth,
height=\figureheight,
area legend,
scale only axis,
xmin=731000,
xmax=735500,
xtick={731307,731672,732038,732403,732770,733135,733501,733866,734233,734598,734964},
xticklabels={2002,2003,2004,2005,2006,2007,2008,2009,2010,2011,2012},
ymin=0,
ymax=700,
title={Historical cash flows (Fast storage)},
legend style={draw=black,fill=white,legend cell align=left}
]
\addplot[ybar,bar width=0.0185396825396825\figurewidth,bar shift=-0.0115873015873016\figurewidth,fill=mycolor1,draw=black] plot coordinates{(731307,209.55721839653)
(731672,153.518311613648)
(732038,325.773520865301)
(732403,458.942753812938)
(732770,590.561488508495)
(733135,395.977992727299)
(733501,410.109964524883)
(733866,286.985956561413)
(734233,250.009927013988)
(734598,124.574766939602)
(734964,204.659754265259)};

\addlegendentry{SpotModel1};

\addplot [
color=black,
solid,
forget plot
]
table[row sep=crcr]{
731000 0\\
735500 0\\
};
\addplot[ybar,bar width=0.0185396825396825\figurewidth,bar shift=0.0115873015873016\figurewidth,fill=red!50!black,draw=black] plot coordinates{(731307,232.692044616128)
(731672,182.57701398651)
(732038,289.401372392679)
(732403,532.109796243799)
(732770,627.029175986714)
(733135,409.841113504705)
(733501,411.907572543964)
(733866,307.519957651918)
(734233,261.842561991553)
(734598,134.859434289143)
(734964,203.079224858682)};

\addlegendentry{SpotModel2};

\end{axis}
\end{tikzpicture}%
\caption{Historical cash flows for spot models 1 and 2 (fast storage)} 
\label{fig:CompareSpotModelsFastStorage} 
\end{figure}

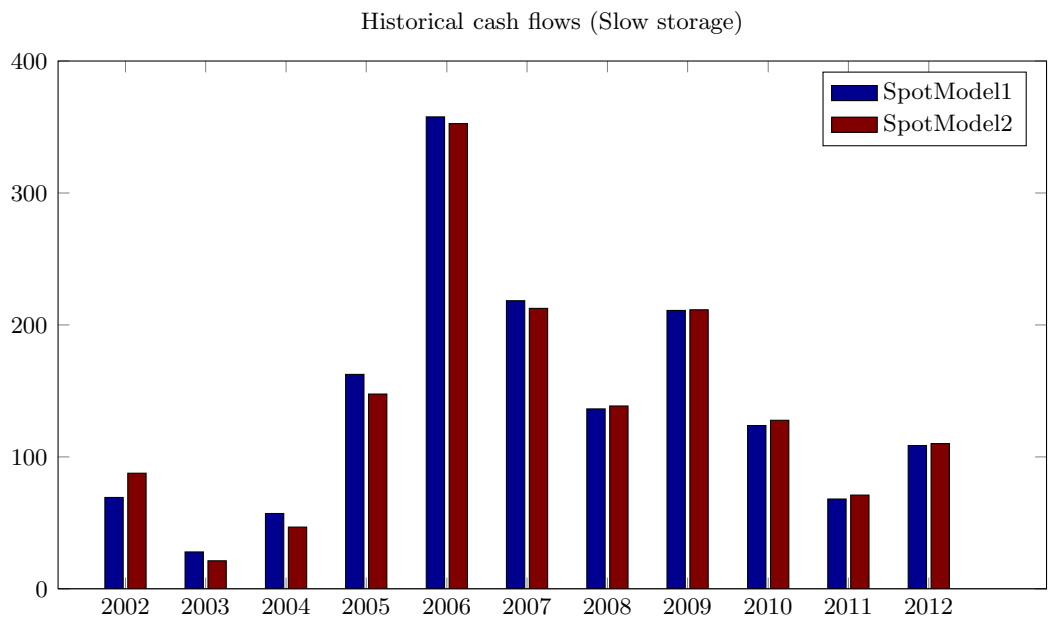
\begin{figure} 
\setlength\figureheight{7cm} 
\setlength\figurewidth{13cm}
%
%
%
%

\definecolor{mycolor1}{rgb}{0,0,0.5625}%

\begin{tikzpicture}

\begin{axis}[%
width=\figurewidth,
height=\figureheight,
area legend,
scale only axis,
xmin=731000,
xmax=735500,
xtick={731307,731672,732038,732403,732770,733135,733501,733866,734233,734598,734964},
xticklabels={2002,2003,2004,2005,2006,2007,2008,2009,2010,2011,2012},
ymin=0,
ymax=400,
title={Historical cash flows (Slow storage)},
legend style={draw=black,fill=white,legend cell align=left}
]
\addplot[ybar,bar width=0.0185396825396825\figurewidth,bar shift=-0.0115873015873016\figurewidth,fill=mycolor1,draw=black] plot coordinates{(731307,69.2895150950574)
(731672,27.998648981782)
(732038,57.1394149545621)
(732403,162.542141783608)
(732770,357.667594849155)
(733135,218.367385546609)
(733501,136.38864583265)
(733866,211.00493260917)
(734233,123.786865954706)
(734598,68.1086954489719)
(734964,108.608666591455)};

\addlegendentry{SpotModel1};

\addplot [
color=black,
solid,
forget plot
]
table[row sep=crcr]{
731000 0\\
735500 0\\
};
\addplot[ybar,bar width=0.0185396825396825\figurewidth,bar shift=0.0115873015873016\figurewidth,fill=red!50!black,draw=black] plot coordinates{(731307,87.6648131960642)
(731672,21.2945965370229)
(732038,46.8342879324451)
(732403,147.65006413124)
(732770,352.573939051781)
(733135,212.565383188692)
(733501,138.601064004879)
(733866,211.504468039404)
(734233,127.780091037968)
(734598,71.0871410317516)
(734964,110.09658295181)};

\addlegendentry{SpotModel2};

\end{axis}
\end{tikzpicture}%
\caption{Historical cash flows for spot models 1 and 2 (slow storage)} 
\label{fig:CompareSpotModelsSlowStorage} 
\end{figure}

We run the valuation procedure explained in Section \ref{SNumRes} for
 the two spot models, during the testing periods between 2003 and 2012, and we describe the performance of both models through historical spot paths: in particular,  we report in Figures \ref{fig:CompareSpotModelsFastStorage} and \ref{fig:CompareSpotModelsSlowStorage} the cumulative cash flows using the optimal spot strategy for historical spot trajectories.
 In the fast storage case, Figure \ref{fig:CompareSpotModelsFastStorage} shows that spot-prompt spread model 2 yields slightly better results than spot model 1 in all the test cases, except for the year 2004.
 In the slow storage case, see Figure \ref{fig:CompareSpotModelsSlowStorage}, the two spot models give comparable results for all periods. In the fast storage case, other tests show that in spot model 2, the cumulative cash flows  generate
a lower standard deviation than in spot model 1, which is in agreement with its better performance.\\

\subsubsection{Effect of spikes modeling}

The presence of spikes in natural gas prices is an essential feature of the
 dynamics of spot prices. In fact, as we noted in Section \ref{stylFact}, these  jumps are sudden dislocations of prices between spot and prompt contracts, due to unpredicted weather changes, technical problems in the transport chain or simply to poor anticipation of the storage capacities of the market. \\

These spikes can be a source of value for the storage manager, since a large gap between spot and prompt prices can be monetized by buying gas during a negative spike, and selling gas during a positive spike. Since these are rapidly absorbed by the market, the value that can be captured from them
strongly depends on storage characteristics. In fact, numerical tests 
show that with slow storage, the spike modeling has less effect
 on the extrinsic value, compared to fast storage. For this reason
 we only concentrate  on the latter.\\

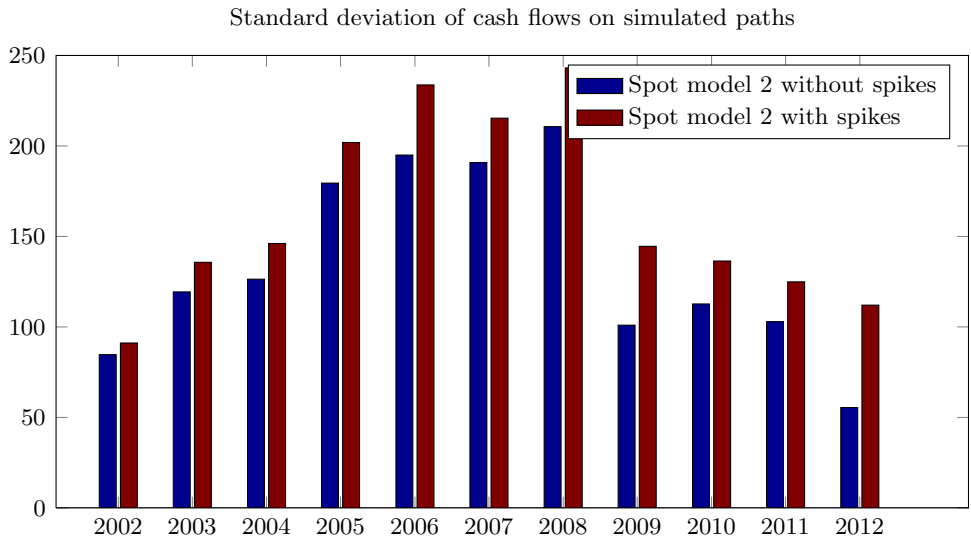
\begin{figure} 
\setlength\figureheight{6cm} 
\setlength\figurewidth{12cm}
%
%
%
%

\definecolor{mycolor1}{rgb}{0,0,0.5625}%

\begin{tikzpicture}

\begin{axis}[%
width=\figurewidth,
height=\figureheight,
area legend,
scale only axis,
xmin=731000,
xmax=735500,
xtick={731307,731672,732038,732403,732770,733135,733501,733866,734233,734598,734964},
xticklabels={2002,2003,2004,2005,2006,2007,2008,2009,2010,2011,2012},
ymin=0,
ymax=250,
title={Standard deviation of cash flows on simulated paths},
legend style={draw=black,fill=white,legend cell align=left}
]
\addplot[ybar,bar width=0.0185396825396825\figurewidth,bar shift=-0.0115873015873016\figurewidth,fill=mycolor1,draw=black] plot coordinates{(731307,84.6937222013084)
(731672,119.360580627104)
(732038,126.376328053432)
(732403,179.479226453656)
(732770,194.969295016291)
(733135,190.85456378652)
(733501,210.681697219038)
(733866,100.975345610834)
(734233,112.698932168278)
(734598,102.931247866366)
(734964,55.4645017363264)};

\addlegendentry{Spot model 2 without spikes};

\addplot [
color=black,
solid,
forget plot
]
table[row sep=crcr]{
731000 0\\
735500 0\\
};
\addplot[ybar,bar width=0.0185396825396825\figurewidth,bar shift=0.0115873015873016\figurewidth,fill=red!50!black,draw=black] plot coordinates{(731307,91.1204423283412)
(731672,135.70826761754)
(732038,146.099744243023)
(732403,201.919776809952)
(732770,233.714820029514)
(733135,215.347863811356)
(733501,243.106232343953)
(733866,144.540437285808)
(734233,136.425584031137)
(734598,124.90877107315)
(734964,112.045629859494)};

\addlegendentry{Spot model 2 with spikes};

\end{axis}
\end{tikzpicture}%
\caption{Standard deviation of cash flows on simulated paths} 
\label{fig:STDFastUnit} 
\end{figure}

\begin{figure} 
\setlength\figureheight{6cm} 
\setlength\figurewidth{12cm}
%
%
%
%

\definecolor{mycolor1}{rgb}{0,0,0.5625}%

\begin{tikzpicture}

\begin{axis}[%
width=\figurewidth,
height=\figureheight,
area legend,
scale only axis,
xmin=731000,
xmax=735500,
xtick={731307,731672,732038,732403,732770,733135,733501,733866,734233,734598,734964},
xticklabels={2002,2003,2004,2005,2006,2007,2008,2009,2010,2011,2012},
ymin=0,
ymax=1000,
title={Expected cash flows on simulated paths},
legend style={draw=black,fill=white,legend cell align=left}
]
\addplot[ybar,bar width=0.0185396825396825\figurewidth,bar shift=-0.0115873015873016\figurewidth,fill=mycolor1,draw=black] plot coordinates{(731307,261.56866829058)
(731672,337.727583687911)
(732038,395.619769672787)
(732403,592.085372307067)
(732770,860.97135810009)
(733135,697.000264086102)
(733501,674.674511820533)
(733866,459.553081648041)
(734233,420.225018343475)
(734598,352.778489555023)
(734964,272.537645440728)};

\addlegendentry{Spot model 2 without spikes};

\addplot [
color=black,
solid,
forget plot
]
table[row sep=crcr]{
731000 0\\
735500 0\\
};
\addplot[ybar,bar width=0.0185396825396825\figurewidth,bar shift=0.0115873015873016\figurewidth,fill=red!50!black,draw=black] plot coordinates{(731307,290.3042092599)
(731672,393.209433531341)
(732038,460.582343633201)
(732403,678.238134938257)
(732770,960.17838627747)
(733135,788.531009537113)
(733501,795.634555715179)
(733866,545.301150648146)
(734233,477.996968243322)
(734598,409.775184294196)
(734964,362.00830299022)};

\addlegendentry{Spot model 2 with spikes};

\end{axis}
\end{tikzpicture}%
\caption{Expected cash flows on simulated paths} 
\label{fig:SpotValueFastUnitSim} 
\end{figure}
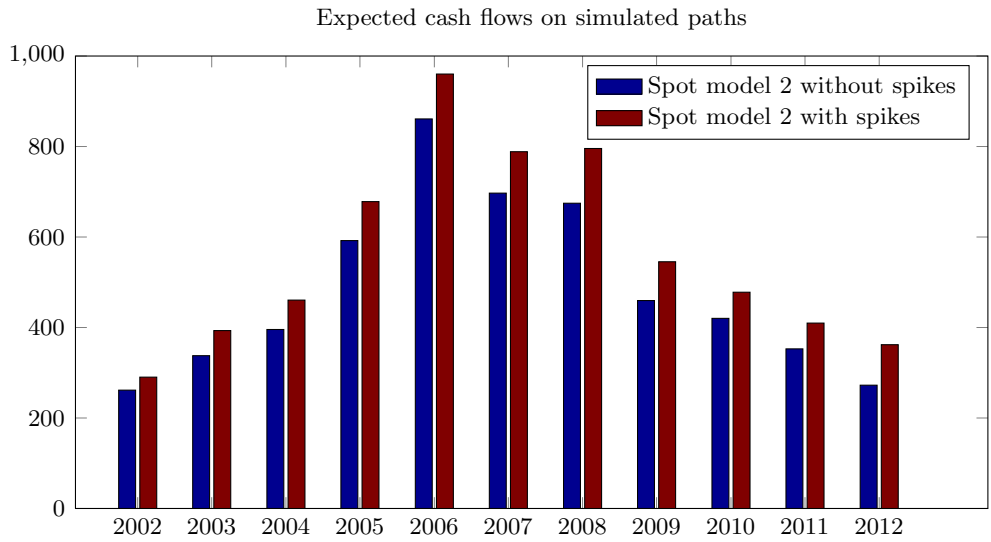

Figure \ref{fig:SpotValueFastUnitSim} represents the expected cumulative cash flows, on simulated paths under the spot model 
\eqref{spotModel2}, for fast storage. All the test periods show 
that modeling the spikes in the spot dynamics gives a larger extrinsic 
value for the storage unit, but at the same time it introduces  
 a larger standard deviation for the cumulative cash flows, as illustrated in
  Figure \ref{fig:STDFastUnit}.  This foresees
 a more significant  uncertainty for the cash flows  on a single sample
 path, when the spikes are taken into account in the spot modeling.\\

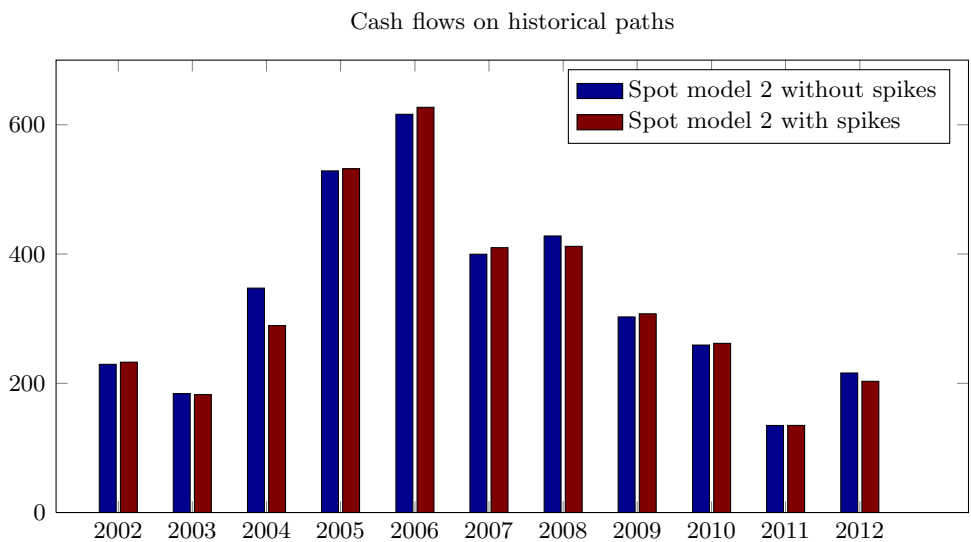
\begin{figure} 
\setlength\figureheight{6cm} 
\setlength\figurewidth{12cm}
%
%
%
%

\definecolor{mycolor1}{rgb}{0,0,0.5625}%

\begin{tikzpicture}

\begin{axis}[%
width=\figurewidth,
height=\figureheight,
area legend,
scale only axis,
xmin=731000,
xmax=735500,
xtick={731307,731672,732038,732403,732770,733135,733501,733866,734233,734598,734964},
xticklabels={2002,2003,2004,2005,2006,2007,2008,2009,2010,2011,2012},
ymin=0,
ymax=700,
title={Cash flows on historical paths},
legend style={draw=black,fill=white,legend cell align=left}
]
\addplot[ybar,bar width=0.0185396825396825\figurewidth,bar shift=-0.0115873015873016\figurewidth,fill=mycolor1,draw=black] plot coordinates{(731307,229.393184237188)
(731672,184.117771743112)
(732038,347.273567712564)
(732403,528.651039310825)
(732770,616.235661783041)
(733135,399.73465546153)
(733501,427.965188601696)
(733866,302.695833772874)
(734233,259.177610669251)
(734598,134.779440111558)
(734964,215.959142806215)};

\addlegendentry{Spot model 2 without spikes};

\addplot [
color=black,
solid,
forget plot
]
table[row sep=crcr]{
731000 0\\
735500 0\\
};
\addplot[ybar,bar width=0.0185396825396825\figurewidth,bar shift=0.0115873015873016\figurewidth,fill=red!50!black,draw=black] plot coordinates{(731307,232.692044616128)
(731672,182.57701398651)
(732038,289.401372392679)
(732403,532.109796243799)
(732770,627.029175986714)
(733135,409.841113504705)
(733501,411.907572543964)
(733866,307.519957651918)
(734233,261.842561991553)
(734598,134.859434289143)
(734964,203.079224858682)};

\addlegendentry{Spot model 2 with spikes};

\end{axis}
\end{tikzpicture}%
\caption{Cash flows on historical paths} 
\label{fig:SpotValueFastUnitHist} 
\end{figure}

A final test of the effect of the spikes modeling is performed on historical spot paths for each test period, and results are shown in Figure \ref{fig:SpotValueFastUnitHist}. 
In fact, according to this graphic, it seems that modeling  the spikes does not make a relevant contribution. This accords with the fact that the models with spikes produce a large standard deviation.
Finally this historical back testing does not show significant advantages of spike modeling. \\

\subsection{Model risk measure}
\label{S72}

In order to quantify the modeling uncertainty, as anticipated, we will consider an approach introduced 
by \cite{cont2006model}, a method that was proposed for the uncertainty of stock models in view of pricing exotic derivative products. In that case, the market data are a set of vanilla option prices (or bid/ask intervals).
 Then the model uncertainty for an exotic payoff $H$, is quantified by computing the range of prices of this exotic product, using a set of risk neutral models $\Gamma$ that calibrate the reference vanilla prices, i.e.
\begin{eqnarray}
\pi(H) = \max_{Q \in \Gamma} \E^Q[H] - \min_{Q \in \Gamma} \E^Q[H].
\label{riskMeasure}
\end{eqnarray}

For our gas storage valuation problem, we will adapt this risk measure, by 
using as \enquote{calibration} data, the historical prices of the  futures and 
spot contracts. The constraint of calibration on vanilla prices is
 replaced by the success of suitable statistical tests and closeness to the
 optimal likelihood objective function value of the model.
 Indeed, in our case, the  family $\Gamma$ consists of a set of spot models, 
\eqref{spotModel1} or \eqref{spotModel2}, which pass statistical tests 
imposed  by the modeling hypothesis for  the noise $(\epsilon_t)$
to be of type Garch(1,1), and have a likelihood function value close
 to the optimal one found during the model estimation.
 This methodology for the generation of the models set $\Gamma$ is partially 
 similar to the one proposed in the case of multi-asset options
 by \cite{lunven2006}. In this study, the authors calibrate a multi-assets 
model to single-asset vanilla options, then build the set $\Gamma$ by
 perturbation of the correlation matrix. This yields a family of models
 that fit perfectly  the reference vanilla options, but
 differ by their correlation matrix.\\

In our case of gas storage valuation, the statistical estimation of the spot model parameters, in 
\eqref{spotModel1} or \eqref{spotModel2}, is realized by classical maximum likelihood methods. The estimation procedure is a maximization problem 
\begin{eqnarray*}
\max_{\theta = \{a_1, a_2, a_3, \kappa, \gamma_1, \alpha_1\}} L(\theta),
\end{eqnarray*}
where $L(\theta)$ is the likelihood function associated with the spot
 model \eqref{spotModel1} or \eqref{spotModel2}.
This maximization yields an optimal parameters vector $\theta^{\star} = \{a_1^{\star}, a_2^{\star}, a_3^{\star}, \kappa^{\star}, \gamma_1^{\star}, \alpha_1^{\star} \}$, an optimal likelihood function value $L(\theta^{\star})$, and an empirical variance-covariance matrix
 $\Sigma^{\star}$ associated with model parameters, which specifies the confidence interval for the estimated parameters, up to a confidence level. \\

In order to generate the spot models family, we perturb the optimal parameters $\theta^{\star}$ by adding a Gaussian noise with the specified covariance matrix $\Sigma^{\star}$ to 
 $\theta^{\star}$. This yields a set of perturbed parameters
 $\{\theta_i\}_{i\in I}$, from which we retain only the perturbed models
that satisfy two constraints:
first,  the inferred Garch white noise  $z(\theta_i)$ in \eqref{garch} 
 has to pass a statistical test of normality \footnote{We use a Kolmogorov-Smirnov test for the normality test of the inferred noise $z$.}; second,
 the corresponding likelihood function value $L(\theta_i)$ 
 has to be  close to the optimal value
 $L(\theta^{\star})$, by a small factor $\epsilon$ i.e.
 $L(\theta_i) > (1-\epsilon) L(\theta^{\star})$.\\

In our framework, $\Gamma$ will be the set of $\theta_i, i \in I$,
fulfilling the two conditions above.


After having constructed the models set $\Gamma$, we can now define
  the associated model risk. 
In our storage valuation problem, the analogue risk measure to \eqref{riskMeasure} 
can be expressed using the value function $J^{\star}(\theta)$ in \eqref{maxPb}, 
where we emphasize the dependence of this value function 
with respect to the spot model parameters $\theta$, and we express the risk measure in relative terms. We set
\begin{eqnarray}
\pi_1 = \frac{\max_{\theta_i \in \Gamma} J^{\star}(\theta_i) - \min_{\theta_i \in \Gamma} J^{\star}(\theta_i)}{J^{\star}(\theta^{\star})}.
\label{riskMeasureGas1}
\end{eqnarray}
In this risk measure evaluation, each $J^{\star}(\theta_i)$,
is calculated using
spot and futures paths simulated under the perturbed model $\theta_i$.\\

Moreover, we propose a second model risk measure based on the
performance on realized historical spot and futures paths. For this we define
\begin{eqnarray}
\pi_2 = \frac{\max_{\theta_i \in \Gamma} \text{Wealth}_{\text{spot+futures}}(\theta_i) - \min_{\theta_i \in \Gamma} \text{Wealth}_{\text{spot+futures}}(\theta_i)}{\text{Wealth}_{\text{spot+futures}}(\theta^{\star})},
\label{riskMeasureGas2}
\end{eqnarray}
where
  $\text{Wealth}_{\text{spot+futures}}$ 
represents the cumulative cash flows,
 computed on the historical path, as defined in \eqref{cashFlowHedg}. \\

The two risk measures $\pi_1$ and $\pi_2$ are computed for each of 
the test periods from 2003 to 2012, under the two spot models 1 and 2, using 
a  set of 30 perturbed models.
 The results reported in  Table \ref{resultsRM} again show a better
 performance for spot model 2. In fact, this model seems
 to be less subject to model risk, since it gives a smaller
 range of prices, compared to spot model 1.

\begin{table}[H]
   \centering
	\begin{tabular}{|c|c|c|c|c|c|c|}
		\hline
		 &\multicolumn{2}{|c|}{Risk measure $\pi_1$} & \multicolumn{2}{|c|}{Risk measure $\pi_2$} \\
		\hline
		Starting date & Spot model 1 & Spot model 2 & Spot model 1 & Spot model 2 \\
		\hline
2003-Apr &51.33 \% &44.8085 \% &70.8465 \% &39.3852 \% \\
2004-Apr &25.4987 \% &23.6942 \% &26.5597 \% &22.3195 \% \\
2005-Apr &26.0388 \% &27.0318 \% &50.7306 \% &38.352 \% \\
2006-Apr &14.9666 \% &15.9873 \% &10.6853 \% &6.6954 \% \\
2007-Apr &93.8336 \% &14.7645 \% &29.4626 \% &18.6143 \% \\
2008-Apr &37.9839 \% &13.8195 \% &16.6811 \% &8.6166 \% \\
2009-Apr &20.7969 \% &10.1216 \% &15.1415 \% &8.1936 \% \\
2010-Apr &26.7845 \% &12.8976 \% &33.0669 \% &7.5285 \% \\
2011-Apr &25.9442 \% &12.3857 \% &35.8704 \% &30.9282 \% \\
2012-Apr &16.7783 \% &9.1489 \% &13.1014 \% &7.1694 \% \\
		\hline
	\end{tabular}
	\caption {\label{resultsRM} Model risk measure for spot models 1 and 2.}
\end{table}

One observation that follows clearly from Table \ref{resultsRM} is that
 the  range of prices  induced by the model uncertainty 
 and measured by $\pi_1$ and $\pi_2$ 
 represents  a large proportion of  the storage value. 
This shows  that the dependence  of gas storage valuation 
 on spot modeling is quite significant.
 While the literature has concentrated its efforts until now 
on the specification of
 an optimal valuation strategy, we believe that one should pay more attention to the choice of spot-futures modeling framework. 
A second comment that we can infer from  Table \ref{resultsRM} is 
that model 2 appears to be less sensitive to the change of parameters and is 
therefore more robust. Fortunately, this is in concordance with the
 better performance of spot model 2 already observed in 
Section \ref{S71}. Table \ref{resultsRM} shows that the spot-futures valuation framework is subject to a large model risk (average: $25\%$). For comparison, 
the model risk for a basket option has been evaluated to $3\%$ (see 
\cite{lunven2006}).

\section{Conclusion}

In this paper we consider the problem of gas storage valuation. After
restating  the main stylized facts of natural gas prices, 
specifically seasonality and spikes, we present a joint modeling framework for the futures curve and the spot, with two different   spot models.
 Using a Monte Carlo simulation method, we estimate 
 the extrinsic optimal spot strategy; for the purpose of variance
reduction of the cumulative cash flow, we set up 
a financial hedging strategy.
We also conduct back testing using historical data of futures and spot
 prices over a period of 10 years. This demonstrates the better
 performance of the extrinsic strategy compared to the
classic intrinsic  futures-based strategy. 
 In fact, the spot strategy allows the manager to better track
  the value of gas storage, in real market conditions,
 even in the case of a slow storage unit.

In the final section, we study the model uncertainty and its effect on
 storage value,  concentrating on the risk associated
with spot modeling. After a quantitative comparison of 
the two spot models we proposed, we conclude that the model based on the spot-prompt spread performs better.
 In order to quantify the stability of these results
 with respect to  model uncertainty, we define two model risk measures, 
inspired by the work of \cite{cont2006model}, but
based on historical prices.
Using those risk measures, we observe the great sensitivity of gas 
storage value with respect to the modeling assumptions. 
In fact the model uncertainty, as measured by the  size of price range,  represents a large proportion of storage value. 
This puts into perspective the concentration of effort in the literature on the specification of an optimal valuation strategy. In fact, much more attention 
should  be probably  devoted to the discussion 
of modeling  assumptions.

\newpage
\begin{appendices}

\section{Different types of gas storage facilities}
\label{appA}

Natural gas storage units are underground facilities, so their characteristics depend essentially on the geological properties of the storage area. There are three types of gas storage units: depleted gas/oil fields, aquifers and salt caverns. The main characteristics that distinguish these gas storage units are their injection/withdrawal rates, the total capacity and the so-called cushion volume and working volume. The cushion volume is the quantity of gas that must remain in the storage unit to provide the required pressurization, and the working volume is the volume of gas that can be extracted. Using the notations in this article, the cushion volume corresponds to the minimum volume $V_{min}$, the total capacity corresponds to $V_{max}$, and the working volume is represented by the actual volume minus the cushion volume, i.e. $V_t-V_{min}$.\\

These characteristics distinguish two different types of gas storage: base-load and peak-load. Base-load units are used to meet seasonal demand (a
 more or less predictable phenomenon). In fact the demand for gas is highly concentrated in the winter season, so in order to ensure sufficient supply, gas is bought and stored in the summer season then withdrawn and sold in  winter. The main characteristics of base-load units are their large volume capacity and
 low deliverability rates.

On the other hand, peak-load units are used to mitigate the risk of unpredictable increases in the gas demand, generally caused by weather changes
 or technical problems in the pipeline system. Hence, they have to be very reactive and have high deliverability rates, higher injection/withdrawal rates, and in general they contain less gas than base-load units.
 While the filling/withdrawal cycle duration of a 
base-load is in general one year, peak-loads
 can have  a turn-over period of a few weeks.\\

The depleted gas/oil fields and aquifers are of the base-load type, while salt caverns are peak-load facility. We summarize here their main characteristics.

\begin{itemize}
\item \textbf{Depleted gas and oil fields}. They are the most commonly used underground storage sites because of their wide availability. Besides their large capacities, they benefit from the already available wells and injection/withdrawal equipments, pipelines etc. Their main drawbacks are low deliverability rate and the large \textit{cushion} gas percentage (although part of this non-usable gas already  exists in the geological formation). Therefore, these depleted fields naturally belong to the base-load category.


\item \textbf{Aquifers}. These are underground, porous and permeable rock formations that act as natural water reservoirs.

They are flexible units with small volume, but more expensive than depleted fields since everything has to be built from scratch (wells, extraction equipments, pipelines, etc).  In some instances, the  installment of
aquifer infrastructure  can take four years, which is more than twice the time needed to transform depleted reservoirs into storage facilities.

On the other hand, aquifers require a greater percentage of \textit{cushion} gas than depleted reservoirs: up to $80\%$ of the total gas volume.

Like depleted fields, aquifers operate on a single annual cycle, so they still belong to the base-load category.

\item \textbf{Salt cavern}. Salt caverns are the third common choice for gas storage. They are created by dissolving and extracting a  certain amount of salt from the geological formation; this process then leaves a cavern that can be used for natural gas storage.

A salt cavern offers storage with high deliverability, with low \textit{cushion} gas requirements ($30\%$ of \textit{cushion} gas), but with lower
 capacities than depleted fields and aquifers. They cannot be used to meet base-load storage requirements, but they are
well suited to rapid actions, which are distinctive features 
of the peak-load category.\\
\end{itemize}

Table \ref{storageTypes}, compiled by the Federal Energy Regulatory Commission (FERC),\footnote{\textit{Current State of and Issues Concerning Underground Natural Gas Storage}, Federal Energy Regulatory Commission (FERC), Staff Report, September 30, 2004. cf: \url{http://www.ferc.gov/EventCalendar/Files/20041020081349-final-gs-report.pdf}} summarizes the three types of storage and their characteristics. 
\begin{table}[H]
   \centering
	\begin{tabular}{|p{3cm}||p{4cm}||p{3cm}||p{4cm}|}
		\hline
		 \textbf{Type} & \textbf{Cushion to working gas ratio} & \textbf{Injection period (days)} & \textbf{Withdrawal period (days)} \\
		\hline
		Aquifer & Cushion $50\%$ to $80\%$ & 200 to 250 & 100 to 150\\
		\hline
		Depleted oil/gas reservoirs & Cushion $50\%$ & 200 to 250 & 100 to 150\\
		\hline
		Salt cavern & Cushion $20\%$ to $30\%$ & 20 to 40 & 10 to 20\\
		\hline
	\end{tabular}
	\caption {\label{storageTypes} Types of natural gas storage.}
\end{table}


\newpage
\section{Futures-based valuation methods}
\label{appB}

Futures-based valuation is still  very commonly used in natural gas storage management. This is mainly due to its simplicity and its low risk profile. It is based on trading  natural gas futures with physical delivery, combined with gas injection/withdrawal. The idea is the following: at the beginning of the storage contract, the manager observes the initial futures curve, and decides to buy/sell multiple forward contracts, and consequently receives/delivers natural gas at their expiration, during a delivery period.\footnote{For example, the Nymex NG futures have monthly spaced maturities, and 
the delivery period spreads out over the month following each maturity date.}\\

In order to determine the optimal futures positions, a linear optimization problem has to be  solved, with constraints imposed by the physical conditions of the storage unit, cf \cite{eydelandBook}.
Let us denote by $(F(t,T_j))_{j=1,...,N}$ the available futures contracts in the markets (the maturities $T_j$ being generally spaced monthly), at date $t$, and $\alpha_j(t)$ the quantity of futures $F(t,T_j)$ bought/sold. 
Using the notations introduced in Section \ref{S31}, 
at the initial date $t_0$,
the storage manager chooses futures positions $\alpha_j(t)
\equiv \alpha_j(t_0)$ in order to maximize the profit, under the physical 
constraints fixed by the maximum $V_{max}$  and the  minimum $V_{min}$ storage capacity and the injection $ a_{inj}$ and withdrawal $a_{with}$ rates of the storage unit.\\

A natural optimization problem, at each time $t$, in  the variables
$\alpha_j(t)$,  is the following:
\begin{align}
\label{IntrinsicValue}
\tag{$O_t$}
\begin{split}
IV(t) := \max_{(\alpha_j(t))_{j=1,...,N}} - \Sum_{j} \alpha_j(t) F(t,T_j)   \\
-a_{with} \Delta T_j \leq \alpha_j(t) \leq a_{inj} \Delta T_j \text{,
 for } j=1,...,N \\
V_{min} \leq V(t) + \Sum_{j=1}^{n} \alpha_j(t) \leq V_{max} \text{, for } n=1,...,N, 
\end{split}
\end{align}
where $\Delta T_j = T_{j+1} - T_j$ is equal to the time length of the delivery period of the $F(t,T_j)$ futures and $V(t)$ is the gas volume in the storage at date $t$. \\

We emphasize that the storage manager, who follows the 
intrinsic value methodology, solves $(O_t)$ only at time $t=t_0$.
 $IV(t_0)$ represents the optimal profit 
given by the maximization problem \eqref{IntrinsicValue}; it is generally called {\it intrinsic value}. Indeed the storage manager keeps the optimal 
futures positions $\alpha_j^*(t_0)$ for the whole storage contract duration,
 so this strategy does not take advantage of possible profitable movements of the futures curve. \\

This static methodology was extended by \cite{GrayIntrinsic} to the \textit{rolling intrinsic valuation}, to take advantage of the changing dynamics of the futures curve. We  consider a set of trading dates $t_0<t_1<...<t_{n-1}< t_n  $,
where $t_n$ is the maturity of the storage contract,
at which the manager can buy/sell physical futures.
In the sequel, for $t = t_i, 1 \le i \le n-1$, we set
$\Delta t = t_{i+1} - t_i$.

At the beginning  date $t_0$  of the storage contract,
 optimal futures positions are chosen,  but after the time period 
$\Delta t$, 
 one recalculates the new optimal futures positions: if the manager 
finds it more profitable, the portfolio is rebalanced.
More precisely, suppose that at date $t$, the manager owns
a futures portfolio $\alpha^\star(t)$;
then, at date
 $t+\Delta t$, the manager solves $O_{t+\Delta t}$, calculating
an optimal portfolio $\alpha(t+\Delta t)^{\star}$ and $IV(t+\Delta t)$.
 The profit/cost of rebalancing the futures portfolio from 
$\alpha(t)^{\star}$  to $\alpha(t+\Delta t)^{\star}$ 
 would be equal to
$$
C(t, \Delta t):= \Sum_{j} [\alpha^{\star}_j(t)-\alpha^{\star}_j(t+\Delta t)] F(t+\Delta t,T_j),
$$
so, the manager will switch positions 
only if  this rebalancing profit/cost  $C(t, \Delta t)$
  is positive. 
We denote by
$RI(t)$ the so-called rolling intrinsic value at time $t$.
At time $t_0$ we set $RI(t) = IV(t)$; recursively 
we define  $RI(t+\Delta t):= RI(t) + \max(C(t, \Delta t),0) $; 
$RI(t)$ represents the cumulative profit generated by this enhanced strategy.
Obviously, at each rebalancing date $t$,  the rolling  intrinsic value 
$RI(t)$ is always greater or equal than the
  intrinsic value  $IV(t_0)$. 


\begin{remark}
The intrinsic and rolling intrinsic methodologies capture the
 predictable seasonal pattern of natural gas prices: they  basically lead
  to buying cheap summer futures and selling expensive winter futures. 
 Indeed the obtained storage value greatly depends on the seasonal spread
 between cold and warm periods of the year. The recent tightening of
 seasonal spreads (cf Section \ref{stylFact}) 
implies that
the (rolling) intrinsic value becomes less attractive for practitioners.
\end{remark}

\end{appendices}

\noindent {\bf ACKNOWLEDGEMENTS:} The research of the second
and third named authors  was partially 
supported by the ANR Project MASTERIE 2010 BLAN-0121-01.

\bibliographystyle{plainnat}
\bibliography{biblio}

\end{document}